\documentclass[submission, Phys]{SciPost}

\usepackage{amsmath,braket,mathdots,mathtools,amssymb,xcolor,calligra,color}
\usepackage{graphicx}
\usepackage{hyperref}
\usepackage{lipsum}
\usepackage{bbm}
\def\be{\begin{equation}}
\def\ee{\end{equation}}
\def\bea{\begin{eqnarray}}
\def\eea{\end{eqnarray}}
\def\nn{\nonumber\\}
\def\fr#1{(\ref{#1})}
\begin{document}
\begin{center}{\Large \textbf{
Projective phase measurements in one-dimensional Bose gases
}}\end{center}

\begin{center}
Yuri D. van Nieuwkerk\textsuperscript{1*},
J\"org Schmiedmayer\textsuperscript{2} and
Fabian H.L. Essler\textsuperscript{1}
\end{center}
\begin{center}
{\bf 1} Rudolf Peierls Centre for Theoretical Physics, Parks Road, Oxford OX1 3PU\\
{\bf 2} Vienna Center for Quantum Science and Technology (VCQ), Atominstitut, TU-Wien, Vienna, Austria\\
* yuri.vannieuwkerk@physics.ox.ac.uk
\end{center}

\begin{center}
\today
\end{center}


\section*{Abstract}
{\bf
We consider time-of-flight measurements in split one-dimensional Bose
gases. It is well known that the low-energy sector of such systems can
be described in terms of two compact phase fields $\hat{\phi}_{a,s}(x)$. 
Building on existing results in the literature we discuss
how a single projective measurement of the particle density after trap
release is in a certain limit related to the
eigenvalues of the vertex operator $e^{i\hat{\phi}_a(x)}$. 
We emphasize the theoretical assumptions
underlying the analysis of ``single-shot'' interference
patterns and show that such measurements give direct access to multi-point correlation
functions of $e^{i\hat{\phi}_a(x)}$ in a substantial parameter regime. 
For experimentally relevant situations,
we derive an expression for the measured particle density
after trap release in terms of convolutions of the
eigenvalues of vertex operators involving both sectors of the
two-component Luttinger liquid that describes the low-energy regime of
the split condensate. This opens the door to accessing properties of
the symmetric sector via an appropriate analysis of existing
experimental data.
}

\vspace{10pt}
\noindent\rule{\textwidth}{1pt}
\tableofcontents\thispagestyle{fancy}
\noindent\rule{\textwidth}{1pt}
\vspace{10pt}

\section{Introduction} 
\label{sec:introduction}
The purpose of this manuscript is to revisit the theoretical basis
for the analysis of matter-wave interferometry experiments on split 
one-dimensional Bose gases \cite{Schumm2005,Hofferberth2007,Hofferberth2008,Gring2012,Kuhnert2013,Shin2004,Shin2005,Jo2007,Baumgaertner2010,Corman2014,Aidelsburger2017}. 
In these experiments a trapped (quasi) one-dimensional Bose gas
is first split in two, then allowed to time evolve under an
interacting Hamiltonian, released into three-dimensional space and
finally measured after a given period of free evolution. The
measurement of the particle density after free evolution exhibits
interference fringes. Repeating the experimental sequence many times
provides an enormous amount of information on the quantum mechanical
state of the many-particle system before trap release. Histograms of
the observed interference patterns provide the full quantum mechanical
distribution function of the measured
observable\cite{Gritsev2006,Imambekov2007,Hofferberth2008,Rath2010,Gring2012,Langen2012,Kuhnert2013,AduSmith2013,Schweigler2017}.
The ability of measuring distributions functions of physical
observables in interacting many-particle systems (out of equilibrium)
is a very exciting feature of cold atom experiments
\cite{greiner}, but poses a
formidable theoretical problem and so far only few results have been
obtained in the literature
\cite{cd-07,Imambekov2007,lp-08,ia-13,sk-13,e-13,k-14,sp-17,CoEG17,nr-17,hb-17,lddz-15,bpc-18,gec18}.
In the case of split one-dimensional Bose condensates
the probability distributions of the observed interference patterns
have been analyzed in the framework of Luttinger liquid theory and
very good agreement with experimental observations has been 
found\cite{Altman2004,Bistritzer2007,Burkov2007,Kitagawa2010,Kitagawa2011}. Here
we provide a detailed derivation for the fit formula used to analyze the
experimental data \cite{Kuhnert2013} for individual
measurements. The formula is obtained in a particular
limit of a new theoretical expression that describes projective
density measurements in time of flight experiments.
Like previous work our approach is based on the Luttinger liquid
description of the  phase degrees of freedom. We discuss why this
analysis is restricted to the weakly interacting regime, and what
modifications emerge for stronger interactions. Our derivation makes
it clear why such measurements provide access to equal time
multi-point correlation functions of vertex operators of the phase field.  

This paper is organized as follows: in Section
\ref{sec:setup_and_time_of_flight}, we review the setup for
time-of-flight experiments and how measured properties are related
to quantities in the split gases before trap release.
In Section \ref{sec:measured_observable}, we express the measured
density after time-of-flight in terms of an appropriate vertex
operator in the field theory describing the low-energy
degrees of freedom of the one-dimensional gas. Section
\ref{sec:eigenstates_of_the_vertex_operator} shows how to construct a
basis of eigenstates for these operators. In Section
\ref{sec:application_vertex_operator_eigenstates_for_coherently_split_bose_gases},
we show that the experiments can be viewed as projective measurements
that sample the eigenvalues of the vertex operator according to a
probability distribution that is fixed by the state which the system is
initialized in after the splitting procedure.

As an example, we consider the case of coherently split bose gases
without tunnel coupling, \emph{cf.} Refs. \cite{Kitagawa2010,Kitagawa2011}. 


\section{Setup and time-of-flight recombination} 
\label{sec:setup_and_time_of_flight}

We consider a pair of one-dimensional bose gases of length $L$. We
denote the longitudinal (along the 1D direction) and transverse
coordinates by $x$ and $\vec{r}$ respectively. The corresponding
momentum coordinates will be denoted by $(k,\vec{p})$ and we use units such that $\hbar = 1$ throughout the paper. The gases are
placed at transverse positions $\vec{r}_{1,2} = \pm \vec{d}/2$. In the
first stage of the experiment, the two condensates time-evolve under
some one-dimensional Hamiltonian $H_{\mathrm{1d}}$, until a time $t_{0}$. In the second
stage, they are released from the 
trap, causing them to expand in three-dimensional space and overlap. Finally, the three-dimensional gas density is
measured after a ``time of flight'' $t_{1}$. We model this measurement
by assuming that the many-particle wave function collapses to a
simultaneous eigenstate $|\Psi\rangle$ of the operators
\begin{align}
\hat{\rho}_{\mathrm{tof}}(x, \vec{r}, t_{1} + t_{0}) =
\hat{\Psi}^{\dagger}(x, \vec{r}, t_{1} +
t_{0})\hat{\Psi}(x, \vec{r}, t_{1} + t_{0}),
\end{align}
where $\hat{\Psi}_{\mathrm{tof}}(x,\vec{r},t)$ are Heisenberg picture boson
annihilation operators at position $(x,\vec{r})$ and time $t$. They
satisfy equal-time commutation relations
\begin{align}
\left[\hat{\Psi}(x, \vec{r}),\hat{\Psi}^{\dagger}(z, \vec{r}^{\prime})\right] = \delta(x-z)\delta^{2}(\vec{r}-\vec{r}^{\prime}),
\end{align}
with all other commutators being zero. Importantly the density
operators $\hat{\rho}_{\mathrm{tof}}(x, \vec{r}, t_{1} + t_{0})$ at
different positions commute. This implies that the measurement outcome
is the function $\varrho_{\mathrm{tof}}(x, \vec{r}, t_{1} + t_{0})$
describing the eigenvalues of the density operators on the
simultaneous eigenstate $|\Psi\rangle$.

We now turn to the relation between $\hat{\Psi}(x,\vec{r}, t)$
and the field operators $\hat{\psi}_{1,2}(x,t_{0})$
describing the two one-dimensional gases at the time $t_{0}$ of the
trap release \cite{Polkovnikov2006,Schaff2015}. We have
\begin{align}
\hat{\Psi}(x, \vec{r}, t) = U^{\dagger}\left(
t;t_0 \right) \hat{\Psi}(x, \vec{r} , t_0) U\left( t;t_0 \right)\ ,
\label{eq:time_evolution_FO_tof_generic}
\end{align}
where $U^{\dagger}(t;t_0) = T \exp i \int_{t_0}^{t_{1}}dt H(t)$ is the
time evolution operator describing the free expansion after the trap
release. This expansion can be analyzed by distinguishing between the ``transverse'' motion, occurring perpendicular to the one-dimensional gas, and the expansion along the one-dimensional gas direction, which is customarily referred to as ``longitudinal''. We retain this nomenclature even though we will impose periodic boundary conditions on the one-dimensional gas for simplicity (see Section \ref{sec:measured_observable}). Open boundary conditions can be accommodated straightforwardly in our approach, but as our focus is on ``bulk'' physics we leave the discussion of boundary effects to future work. We will make two simplifying assumptions \cite{Polkovnikov2006,Schaff2015} about the expansion of the gas after trap release:
\begin{enumerate}
\item{} The state of the gas before its release factorizes into
transverse and longitudinal degrees of freedom. The longitudinal state
is the complicated many-body state we are interested in. The
transverse degrees of freedom occupy the ground state of a harmonic
oscillator potential, with vanishing overlap between the two wells. The
wells are assumed to have a large transverse trapping frequency
$\omega_{\perp}$. This implies that the spatial distribution of the transverse
state is a spatially narrow Gaussian, ensuring that the velocity
distribution in the transverse directions is much broader than in the
longitudinal direction. In some works \cite{Kuhnert2013,Schaff2015} it
is therefore assumed that the longitudinal degrees of freedom are
effectively frozen on the timescales relevant for expansion. Relaxing
this simplifying assumption leads to a more involved description
\cite{Imambekov2009,Manz2010}. In what follows, results based on
frozen longitudinal dynamics will be presented alongside results for
the full, three-dimensional expansion.
\item{} The gases are assumed to evolve as free particles after they
  have been released from the trap. 
  For a justification of this assumption, 
    the reader is referred to \cite{Imambekov2009}.
\end{enumerate}  
Under these assumptions the time evolution after trap release is
described by
\be
U( t;t_0) = e^{- i (t-t_0)
\left(\hat{P}_{x}^{2}+\hat{\vec{P}}_{\perp}^{2}\right)/2m}.
 \label{eq:time_evol_oper_2d}  
\ee
Here $\hat{P}_{x}\, (\hat{\vec{P}}_{\perp})$ is the total momentum
operator in the longitudinal (transverse) direction and $m$ is the 
mass of the individual particles. It is now straightforward to obtain 
the desired relation between the field operators at the time of 
measurement ($t=t_1+t_0$) and the time of trap
release ($t=t_0$),
\begin{align}
\hat{\Psi}_{\mathrm{tof}}(x, \vec{r}, t_1+t_0) = \int
\frac{dk\,d^{2}\vec{p}\,dy\,d^{2}\tilde{\vec{r}}}{\left( 2 \pi
  \right)^{3} } e^{-ik(x-y)}e^{-i \vec{p} \cdot \left( \vec{r} -
  \tilde{\vec{r}} \right) } e^{-i t_1 \frac{k^{2}+\vec{p}^{2}}{2m}}
\hat{\Psi}(y, \tilde{\vec{r}},t_0). 
\label{eq:expansion_time_evol_intermed} 
\end{align}
From our previous discussion we know that at $t=t_0$ a basis of single-particle
states (in the low-energy sector of the Hilbert space) is obtained by
having a boson at position $x$ that is the ground state of one of the
transverse harmonic oscillators centred at $\pm \vec{d}/2$ in the
transverse directions. This implies that the Bose field can be decomposed as 
\begin{align}
\hat{\Psi}(x, \vec{r}, t_0) = \hat{\psi}_{1}(x,t_0)g(\vec{r} +
\vec{d}/2) + \hat{\psi}_{2}(x,t_0)g(\vec{r} -
\vec{d}/2), \label{eq:field_op_two_wells} 
\end{align}
where $\hat{\psi}_{1,2}(x,t_0)$ creates a boson at position $x$ in the ground state
of the transverse harmonic oscillator centred at $\pm \vec{d}/2$ and
$g(\vec{r}\pm \vec{d}/2)$ denotes the corresponding ground state wave functions.
The Bose fields $\hat\psi_i(x,t_0)$ have equal time commutation relations
$[\hat{\psi}_{i}(x,t) , \hat{\psi}^{\dagger}_{j}(z,t)] = 
\delta_{i,j}\delta(x-z)$.
Inserting the decomposition (\ref{eq:field_op_two_wells}) into
(\ref{eq:expansion_time_evol_intermed}), using $g(\vec{x})
\sim e^{- \frac{m \omega}{2} \vec{x}^{2}}$ and assuming that $t_1 \gg 1/
\omega$ (where $\omega$ is the frequency of the harmonic potential in
the transverse direction) then gives
\begin{align}
\hat{\Psi}(x,\vec{r},t_{1}+t_0) = f( \vec{r},t_{1}) \int dy\,
G(x-y,t_{1}) \left[\hat{\psi}_{1}(y,t_0)e^{i \frac{m}{2t_{1}}(\vec{r} +
    \vec{d}/2 )^{2}} + \hat{\psi}_{2}(y,t_0)e^{i \frac{m}{2t_{1}} (\vec{r}
    - \vec{d}/2 )^{2}}  \right], 
\label{eq:final_relation_field_operator_tof} 
\end{align}
where the function $f(\vec{r},t_{1})$ is a Gaussian envelope, and
$G(x,t_{1})$ is a free, single-particle Green's function. The precise
form of these functions, together with the details of the calculation,
are given in Appendix
\ref{sec:relation_between_density_operators_before_and_after_release}. 

Using (\ref{eq:final_relation_field_operator_tof}) we can identify the
observable that is ultimately measured in the time-of-flight
experiments as
\bea
\hat{\rho}_{\mathrm{tof}}(x,\vec{r},t_{1}+t_0)&=&
\left|f(\vec{r},t_{1})\right|^{2} \iint dy\,dz\, G^{*}(x-y,t_{1})
G(x-z,t_{1}) \Big[
  \hat{\psi}_{1}^{\dagger}(y,t_0)\hat{\psi}_{1}(z,t_0)\nn
&+&   \hat{\psi}_{2}^{\dagger}(y)\hat{\psi}_{2}(z)
  + \hat{\psi}_{1}^{\dagger}(y)\hat{\psi}_{2}(z)e^{-i \vec{d}\cdot
    \vec{r}\,m/t_{1}} +
  \hat{\psi}_{2}^{\dagger}(y)\hat{\psi}_{1}(z)e^{i \vec{d}\cdot
    \vec{r}\,m/t_{1}} \Big].
\label{eq:density_tof}
\eea
Each measurement will select one of the eigenvalues of the above sum
of operators. Importantly, the various terms in (\ref{eq:density_tof})
do not commute with one another. Hence at the level of the ``full''
Bose gases the measured observable is not simple.

\subsection{Simplification when the longitudinal expansion is frozen}
Denoting by $\hat{\rho}(t_{0})$ the density matrix of the system at the time of
the trap release, the subsequent evolution is given by
\be
\hat{\rho}(t)=U(t;t_0)\hat{\rho}(t_0)U^\dagger( t;t_0)\ .
\ee
In cases where $\hat{\rho}(t_0)$ and $t_1$ are such that expansion in the
longitudinal direction can be neglected, \emph{cf.} the discussion
above, we have
\be
\hat{\rho}(t_1+t_0)\approx
\widetilde{U}(t_1+t_0;t_0)\hat{\rho}(t_0)\widetilde{U}^\dagger(t_1+t_0;t_0)\ , 
\quad
\widetilde{U}( t_{1}+t_0;t_0) = e^{- i t_{1}\hat{\vec{P}}_{\perp}^{2}/2m}.
\label{special}
\ee
In this case \fr{eq:final_relation_field_operator_tof} can be replaced by
\begin{align}
\hat{\Psi}(x,\vec{r},t_{1}+t_0) = f( \vec{r},t_{1})
\left[\hat{\psi}_{1}(x,t_0)e^{i \frac{m}{2t_{1}}(\vec{r} + \vec{d}/2
    )^{2}} + \hat{\psi}_{2}(x,t_0)e^{i \frac{m}{2t_{1}} (\vec{r} -
    \vec{d}/2 )^{2}} \right] .
\label{specialPsi}
\end{align}
This then results in the following expression for the measured density
\bea
\label{eq:density_tof_noex}
\hat{\rho}_{\mathrm{tof}}(x,\vec{r},t_{1}+t_0)&=&  \left|f(
\vec{r},t_{1})\right|^{2} \Big[\hat{\psi}_{1}^{\dagger}(x,t_0)\hat{\psi}_{1}(x,t_0) +
  \hat{\psi}_{2}^{\dagger}(x,t_0)\hat{\psi}_{2}(x,t_0)\nn
&+& \hat{\psi}_{1}^{\dagger}(x,t_0)\hat{\psi}_{2}(x,t_0)e^{-i \vec{d}\cdot
    \vec{r}\,m/t_{1}} +
  \hat{\psi}_{2}^{\dagger}(x,t_0)\hat{\psi}_{1}(x,t_0)e^{i \vec{d}\cdot
    \vec{r}\,m/t_{1}} \Big]. 
\eea
\section{Luttinger liquid description of the low-energy degrees of freedom}
\label{sec:measured_observable}
We have seen how the field operator after time of flight can be
related to the separate field operators of the original
one-dimensional gases. We focus on the case where the dynamics in
the trap is governed by a Hamiltonian of the form
\begin{align}
H_{\mathrm{1d}} &= \sum_{j=1,2} \int_{-L/2}^{L/2} dx\,\Bigg[
  \frac{1}{2m} \partial_{x} \hat{\psi}_{j}^{\dagger}(x) \partial_{x}
  \hat{\psi}_{j}(x) + g\, \hat{\psi}_{j}^{\dagger}(x) \hat{\psi}_{j}^{\dagger}(x)
  \hat{\psi}_{j}(x) \hat{\psi}_{j}(x) \Bigg] + H_{\mathrm{pert}}.
\label{eq:micr_1d_ham}
\end{align}
We will be interested in cases where $H_{\mathrm{pert}}$ can be considered as a weak perturbation in the sense that it does not change the nature of the low energy degrees of freedom. An example would be a weak tunneling term between the two condensates. 

For ease of exposition, we will assume periodic boundary conditions in the one-dimensional bose gas. This means that coordinates $x = \pm L/2$ are associated with each other during evolution under the Hamiltonian (\ref{eq:micr_1d_ham}). After trap release, these points become independent, and the bosons are supported on all of $\mathbb{R}^{3}$. This somewhat artificial treatment has the advantage that it simplifies our expressions. It must be stressed that a model with open boundary conditions can easily be incorporated into our analysis. Doing so will not, however, change our argument in a fundamental way for regions that are sufficiently far from the edges of the trap.

\subsection{Low energy projection}
\label{sub:low_energy_projection_of_the_density_operator}
In the low-energy sector of the theory dramatic simplifications
occur. The low-energy degrees of freedom can be described by
bosonization \cite{Haldane1981}
\begin{align}
\hat{\psi}_{j}^{\dagger}(x) \sim \sqrt{\rho_{0} + \frac{\partial_{x} \hat{\theta}_{j}(x)}{\pi}}\; e^{-i \hat{\phi}_{j}(x)} \sum_{m} A_{m} e^{2im \left( \hat{\theta}_{j}(x) + \pi  \rho_{0} x\right)}, \;\;\;\;\; j=1,2 . \label{eq:bosonization_identity}
\end{align}
Here the fields $\partial_{x} \hat{\theta}_{j}(x)/\pi$ and
$\hat{\phi}_{j}(x)$ describe long-wavelength fluctuations of density
and phase and have commutation relations
\begin{align}
\left[ \frac{\partial_{x} \hat{\theta}_{i}(x)}{\pi}, \hat{\phi}_{j}(z) \right] = i \delta_{i,j} \delta (x - z). \label{eq:comm1}
\end{align}
The bosonized description applies above a ``cutoff'' that is set by
the healing length $\xi = \pi/mv$ for weakly interacting
bosons, with $v$ the velocity of sound. Bosonizing the Hamiltonian (\ref{eq:micr_1d_ham}) leads to a
perturbed two-component Luttinger liquid of the form (see
Appendix \ref{sec:details_of_the_bosonized_hamiltonian} for details)
\begin{align}
{\cal H} = \sum_{j=s,a} \frac{v}{2 \pi} \int_{-L/2}^{L/2}
dx\; \left[K (\partial_{x} \hat{\phi}_{j}(x))^{2} + \frac{1}{K}
  (\partial_{x} \hat{\theta}_{j}(x))^{2} \right] + {\cal H}_{\mathrm{pert}},
\label{eq:Ham_Bosonized}
\end{align}
where ${\cal H}_{\mathrm{pert}}$ is the low-energy projection of
$H_{\mathrm{pert}}$ and where we have defined symmetric and
antisymmetric combinations of the fields by
\begin{align}
\hat{\phi}_{a} = \hat{\phi}_{1} - \hat{\phi}_{2}, \;\;
\hat{\phi}_{s} = \hat{\phi}_{1} + \hat{\phi}_{2}, \;\;
\hat{\theta}_{a} = \frac{\hat{\theta}_{1} - \hat{\theta}_{2}}{2}, \;\;
\hat{\theta}_{s} = \frac{\hat{\theta}_{1} +
  \hat{\theta}_{2}}{2}.
\label{eq:symm_antisymm_trafo} 
\end{align}
In order for (\ref{eq:Ham_Bosonized}) to apply we require that
${\cal H}_{\mathrm{pert}}$ can be treated as a perturbation in the
sense that it does not invalidate a low-energy description in terms of
phase fields. An example \cite{Gritsev2007,Schweigler2017} is a small tunneling term (with $\lambda\ll mv^2$ proportional to the tunneling amplitude)
\be
H_{\mathrm{pert}}=\lambda\int dx \left[\hat{\psi}^\dagger_1(x)\hat{\psi}_2(x)+{\rm h.c.}\right],
\ee
giving a relevant (in the renormalization group sense)
perturbation of the form 
\be
{\cal H}_{\mathrm{pert}}= \lambda'\int dx\ \cos\hat{\phi}_a(x)\ .
\ee

\subsection{Case with no longitudinal expansion and weak interactions}
We first discuss the simpler case in which the longitudinal expansion
is assumed to be negligible. Applying the bosonization identity
(\ref{eq:bosonization_identity}) to the observable measured in
time-of-flight experiments, the measured density operator 
(\ref{eq:density_tof_noex}) takes the form 
\begin{align}
\hat{\rho}_{\mathrm{tof}}(x,\vec{r},t_{1}+t_0)\simeq 2 \big|f(
\vec{r},&t_{1})\big|^{2} \Bigg\{ |A_0|^2\Big(\rho_{0}
+ \Pi_s(x,t_0)
\Big) \left( 1 + {\rm Re} \left[
  e^{i\hat{\phi}_{a}(x,t_0)+ i\vec{d}\cdot \vec{r}\,\frac{m}{t_{1}}} \right]
\right) \nn
+4 A_{0} A_{1} \Bigg[ \Big(\rho_{0}
+ \Pi_s(x,t_0)
\Big) &\cos \big(2\hat{\theta}_{s}(x,t_0) +2k_{F} x \big) \cos
 2\hat{\theta}_{a}(x,t_0) \bigg[1+ {\rm Re} \big(
  e^{i\hat{\phi}_{a}(x,t_0)+ i\vec{d}\cdot \vec{r}\,\frac{m}{t_{1}}} \big)
  \bigg] \nn
- \Pi_a(x,t_0)
 &\sin\big(2
 \hat{\theta}_{s}(x,t_0) +k_{F} x \big) \sin
 2\hat{\theta}_{a}(x,t_0) \Bigg]+\dots
\Bigg\}, \label{eq:density_tof_bos} 
\end{align}
where we have defined
\be
\Pi_{\alpha}(x,t_0)=\frac{\partial_{x}
  \hat{\theta}_{\alpha}(x,t_0)}{\pi}\ ,\quad \alpha=a,s.
\ee
Here the dots refer to subleading terms in the expansion, in the sense
that the operators have higher scaling dimensions. These operators can
have nonzero expectation values on the states of interest, and they
are multiplied by coefficients $A_{m \neq 0}$. In fact, it has been
shown \cite{Shashi2011} that if $K$ is close to $1$, $A_{0}$
and $A_{1}$ approach each other, and higher order terms cannot simply
be neglected. 

The weakly interacting regime $K \gg 1$ is of particular interest in
view of existing experiments. Here the coefficients
$A_{m \neq 0}$ are small and we need to retain only the first line of
\fr{eq:density_tof_bos}, if the longitudinal expansion during time-of-flight is neglected. This gives
\be
\hat{\rho}_{\mathrm{tof}}(x,\vec{r},t_{1}+t_0)\Bigg|_{K\gg 1}\simeq 2 |A_0|^2\left|f(
\vec{r},t_{1})\right|^{2} \Big(\rho_{0}  + \frac{\partial_{x}
  \hat{\theta}_{s}(x,t_0)}{\pi} \Big) \left( 1 + {\rm Re} \left[
  e^{i\hat{\phi}_{a}(x,t_0)+ i\vec{d}\cdot \vec{r}\, m/t_{1}} \right]
\right).
\label{rhoTOFlargeK}
\ee
As $[\partial_{x} \hat{\theta}_{s}(x,t_0),e^{i\hat{\phi}_{a}(x,t_0)}]=0$,
a projective measurement of $\hat{\rho}_{\mathrm{tof}}$ projects onto
simultaneous eigenstates of these operators.

\subsubsection{Relation of operator eigenvalues to experimental fit
formulas }
\label{sub:relation_between_fit_formula_and_operator_eigenvalues}
In (\ref{rhoTOFlargeK}) the measured density operator has been
expressed as a function of commuting operators
$e^{i\hat{\phi}_{a}(x)}$. A measurement then 
projects onto a simultaneous eigenstate of these operators. Let us
denote the corresponding eigenvalues by the functions
$e^{i\varphi_{a}(x)}$. In the case at hand, i.e. negligible
longitudinal expansion, the density measurement then returns the eigenvalue 
\be
\varrho_{\mathrm{tof}}(x,\vec{r},t_{1}+t_0)\approx 2 \rho_{0} |A_0|^2\left|f(
\vec{r},t_{1})\right|^{2} \left( 1 + {\rm Re} \left[
  e^{i{\varphi_{a}}(x,t_0)+ i\vec{d}\cdot \vec{r}\, m/t_{1}} \right]
\right),
\ee
where it has been assumed that the relevant eigenvalues of
$\partial_x\hat{\theta}_s$ are much smaller than $\rho_{0}$. 
This assumption is justified if the symmetric sector is in 
a thermal state \cite{Kitagawa2011}, where density fluctuations 
are small \cite{Petrov2000}.

In many experiments \cite{Gring2012,Kuhnert2013} the measured gas
density is integrated over a distance $l$ along the longitudinal
coordinate of the gas, giving the measured eigenvalue
\bea
R_{\mathrm{tof}}(\vec{r},t_{1}+t_0,\ell)&=&\int_{-\ell/2}^{\ell/2} dx\ \varrho_{\mathrm{tof}}(x,\vec{r},t_{1}+t_0)\nn
&\approx& 2 \rho_{0}|A_0|^2\left|f( \vec{r},t_{1})\right|^{2} \Big( \ell  +
{\rm Re} \Big[ e^{i\vec{d}\cdot \vec{r}\, m/t_{1}} \int_{-\ell/2}^{\ell/2} dx\, e^{i\varphi_{a}(x,t_0)} \Big] \Big). \label{eq:integrated_density_tof_bos}
\eea
This can now be directly compared to the formula used to fit the
experimentally measured interference fringes given in \cite{Kuhnert2013} as 
\begin{align}
\tilde{R}_{\mathrm{tof}}(\vec{r},t_{1}+t_0,\ell) = 2 \rho_{0} \ell |A_0|^2
\left|f( \vec{r},t_{1})\right|^{2} \left( 1+ C(\ell,t_0) \cos \left( \Phi(\ell,t_0) + \vec{d}\cdot \vec{r}\, m/t_{1} \right) \right). \label{eq:phenom_formula}
\end{align}
Comparing (\ref{eq:phenom_formula}) and
(\ref{eq:integrated_density_tof_bos}) shows that the quantities
$C(\ell)$ and $\Phi(\ell)$ are related to the measured eigenvalues
$e^{i\varphi_{a}(x)}$ by
\begin{align}
C(\ell,t_0) e^{i \Phi(\ell,t_0)} = \frac{1}{\ell} \int_{-\ell/2}^{\ell/2} dx\, e^{i\varphi_{a}(x,t_0)}.
\end{align}

\subsubsection{Determining multipoint correlation functions from
measurements}
\label{sub:multipoint_correlation_functions_of_vertex_operators}
The previous discussion has shown that the experimental measurement of
individual interference patterns permits the determination of the
corresponding vertex-operator eigenvalues $e^{i\varphi_{a}(x)}$. Having
these in hand it is then possible to extract (connected) multi-point
correlation functions from the measurements as
follows \cite{Langen2014,Schweigler2017}. Expectation values
of the form 
\be
g_{\alpha_1,\dots,\alpha_n}(x_{1}, x_{2}, \dots,x_n)
\equiv\langle\psi(t)|\prod_{n} e^{i \alpha_{n} \hat{\phi}(x_{n})}|\psi(t)\rangle
\label{multipoint}
\ee
are obtained by averaging over many measurements of ``single-shot''
interference patterns. According to our previous discussion, each such
measurement provides the eigenvalue $e^{i\varphi_{a}(x)}$ of
$e^{i\hat{\phi}_{a}(x)}$. As vertex operators at different positions
commute with one another, their respective measurements are
independent. Hence the outcome for measuring only
$\prod_{n} e^{i \alpha_{n}  \hat{\phi}(x_{n})}$ is simply given by
the product of the corresponding eigenvalues $\prod_{n} e^{i
  \alpha_{n}  \varphi_{a}(x_{n})}$. These are straightforwardly extracted
from the single-shot measurements discussed above by considering fixed
positions $x_1,\dots,x_n$. Averaging over the outcomes of a large
number of such measurements, and keeping the positions $x_1,\dots,x_n$
fixed throughout provides the desired expectation values \fr{multipoint}.

\subsection{General case in the weakly interacting regime}
We now turn to the case where the longitudinal expansion is not
negligible. In order to have manageable expressions we constrain our
discussion to the regime of weak interactions $K\gg 1$, where we can
set the amplitudes $A_{n\geq 1}=0$. Applying the bosonization identity
(\ref{eq:bosonization_identity}) we then find
\begin{align}
&\hat{\rho}_{\mathrm{tof}}(x,\vec{r},t_{1}+t_0)\simeq 2 \left|f(
\vec{r},t_{1})\right|^{2} |A_0|^2 \iint dy \, dz\, G^{*}(x-y,t_{1})G(x-z,t_{1})\nn
&\times \Bigg\{ \left( \rho_{0} + \frac{\partial_{y}
  \hat{\theta}_{1}(y,t_0) + \partial_{z}
  \hat{\theta}_{1}(z,t_0)}{2 \pi} \right) e^{-i \left( \hat{\phi}_{1}(y,t_0) - \hat{\phi}_{1}(z,t_0) \right) } + \left( 1 \rightarrow 2 \right) \nn
&\qquad+ \left( \rho_{0} + \frac{\partial_{y}
  \hat{\theta}_{1}(y,t_0) + \partial_{z}
  \hat{\theta}_{2}(z,t_0)}{2 \pi} \right) e^{i \left( \hat{\phi}_{1}(z,t_0) - \hat{\phi}_{2}(y,t_0) \right) }e^{i\vec{d}\cdot \vec{r}\, m/t_{1}} + \left( \mathrm{c.c.} \right) \Bigg\} + \ldots. \label{eq:density_tof_bos_longitudinal}
\end{align}
This expression involves products of non-commuting operators, which we
must diagonalize in order to develop a theory of projective
measurements. This significant complication vanishes in the
experimentally relevant case when density fluctuations are small 
compared to the average density $\rho_0$ \cite{Petrov2000}. In that case, the fields
$\partial_{x} \hat{\theta}_{1,2}$ may be neglected, so that the
measured density operator becomes 
\bea
\hat{\rho}_{\mathrm{tof}}(x,\vec{r},t_{1}+t_0)\Bigg|_{K\gg 1}
&\simeq& \rho_{0} \Big| A_{0} f(\vec{r},t_{1}) \int dy\,
G(x-y,t_{1})\Big[ e^{i \frac{m}{2t}\vec{r} \cdot \vec{d}}
  e^{\frac{i}{2}\left( \hat{\phi}_{s}(y,t_0)+ \hat{\phi}_{a}(y,t_0) \right)} \nn
&&\qquad\qquad\qquad + e^{-i    \frac{m}{2t}\vec{r} \cdot \vec{d}} e^{\frac{i}{2}\left(
    \hat{\phi}_{s}(y,t_0) - \hat{\phi}_{a}(y,t_0) \right) }  \Big] \Big|^2 .
\label{eq:density_tof_bos_longitudinal_approx}
\eea
This expression only contains fields which mutually commute. A
measurement thus projects onto simultaneous eigenstates of these
fields, based on some probability distribution which is set by the
state at the time of release. A projective measurement returns the
eigenvalues 
\begin{align}
\varrho_{\mathrm{tof}}(x,\vec{r},t_{1}+t_0)\simeq &\rho_{0} \Big|
A_{0} f(\vec{r},t_{1}) \int dy\, G(x-y,t_{1})\Big[ e^{i \frac{m}{2t}\vec{r} \cdot \vec{d}}
  e^{\frac{i}{2}\left( \varphi_{s}(y,t_0) + \varphi_{a}(y,t_0)
    \right)}\nn
& +  e^{-i \frac{m}{2t}\vec{r} \cdot \vec{d}} e^{\frac{i}{2}\left(
    \varphi_{s}(y,t_0) - \varphi_{a}(y,t_0) \right) }  \Big]
\Big|^2, \label{eq:density_tof_bos_longitudinal_collapsed} 
\end{align} 
where $e^{i\varphi_{a,s}(x,t_0)}$ are the corresponding eigenvalues of
$e^{i\hat{\phi}_{a,s}(x,t_0)}$.

\section{Vertex operator eigenstates}
\label{sec:eigenstates_of_the_vertex_operator}
We now turn to the construction of eigenstates of the vertex operators
$e^{i\hat{\phi}_{a}(x)}$ and corresponding eigenvalues
$e^{i\varphi_{a}(x)}$. The mode expansions for $\hat\phi_a(x)$ and
$\partial_x\hat\theta_a(x)$ are given in Appendix
\ref{sec:details_of_the_bosonized_hamiltonian} and involve zero modes
that reflect the compact nature of the phase fields $\hat\phi_a(x)$.
In particular we have $\hat{\phi}_{a}(x+L) = \hat{\phi}_{a}(x) + 2 \pi
\hat{J}_{a},$ where the eigenvalues $j_a$ of $\hat{J}_{a}$ are integers. We
will consider cases in which the dynamics occurs in the $j_a=0$
subspace, i.e. the initial states lie in this subspace and
$[\hat{J}_a,{\cal H}]=0$. This leaves us with mode expansions of the form
\begin{align}
\hat{\phi}_{a}(x) &= \sum_{j} u_{j} \left( \hat{a}_{j} - \hat{a}_{-j}^{\dagger} \right) e^{i q_{j} x} , \label{eq:phi_vec_form_full} \\
\frac{\partial_{x} \hat{\theta}_{a}(x)}{\pi} &= \frac{-i}{2 u_{0} L} \left( \hat{a}_{0} + \hat{a}_{0}^{\dagger} \right) + \sum_{j \neq 0} \frac{i}{2u_{j}L} \left( \hat{a}_{j} + \hat{a}_{-j}^{\dagger} \right) e^{i q_{j} x}, \label{eq:del_theta_vec_form_full} 
\end{align}
where $q_j=2\pi j/L$, $\Big[ \hat{a}_{j}, \hat{a}^{\dagger}_{k}\Big] = \delta_{j,k}$ and
\begin{align}
u_{j} &= \begin{cases}
	 \Big| \frac{\pi}{2q_{j}LK}\Big|^{1/2}\mathrm{sgn}\left(q_{j}\right), &\text{ for }j \neq 0,\\
	\frac{i}{4} \sqrt{\frac{2v}{K}}&\text{ for }j = 0.
	\end{cases} \label{eq:u_vec_def}
\end{align}
As $[\hat{a}_{k} - \hat{a}_{-k}^{\dagger},\hat{a}_{n} - \hat{a}_{-n}^{\dagger}]=0$ the eigenvalue equation
$e^{i\hat{\phi}_{a}(x)} \ket{\{f_{n}\}} = e^{i\varphi_{a}(x)} \ket{\{f_{n}\}}$
then separates into equations for the individual modes
\begin{align}
u_{k} \left( \hat{a}_{k} - \hat{a}_{-k}^{\dagger} \right) \ket{\{f_{n}\}} = f_{k} \ket{\{f_{n}\}}.
\label{EVeq}
\end{align}
Here the eigenvalues $f_k$ are the Fourier coefficients of the
function $\varphi_{a}(x)$
\be
\varphi_{a}(x)=\sum_{j=0}^\infty f_j\ e^{iq_j x}\ .
\label{eq:f_Fourier}
\ee
As $\hat\phi_a(x)$ is a real field we have $f^{*}_{-n}=f_{n}$ and $f_{0}^{*}=f_{0}$. 
The solution of \fr{EVeq} is
\begin{align}
\ket{\{f_{n}\}}_{a} = \mathcal{N}_{f} \exp \sum_{k} \left( \frac{1}{2} \hat{a}_{k}^{\dagger}\hat{a}_{-k}^{\dagger} + \frac{f_{k}}{u_{k}} \hat{a}_{k}^{\dagger} \right) \ket{0}_{a}, \label{eq:eigenstate}
\end{align}
where $\hat{a}_{k} \ket{0}_{a} = 0$. The normalization constant is
\begin{align}
\mathcal{N}_{f} = \left(\frac{1}{2\pi |u_{0}|^{2}}\right)^{1/4} e^{- \frac{1}{4 |u_{0}|^{2}} f_{0}^{2} } \prod_{k>0} \left(\frac{1}{\pi |u_{k}|^{2}}\right)^{1/2} e^{- \frac{1}{2|u_{k}|^{2}} |f_{k}|^{2}} \label{eq:normalization}
\end{align}
and ensures the normalization of the eigenstates to delta-functions
(see Appendix \ref{sec:normalization_of_vertex_operator_eigenstates}
for details)
\begin{align}
\braket{\{g_{n}\}|\{f_{n}\}}_{a} = \delta(g_{0}-f_{0})\prod_{k > 0}
\delta\big({\rm Re}(g_{k}-f_{k})\big)\delta\big({\rm Im}(g_{k}-f_{k})\big).
\label{eq:delta_normalized}
\end{align}

\section{Application to coherently split Bose gases} 
\label{sec:application_vertex_operator_eigenstates_for_coherently_split_bose_gases}
We now specialize to the case of coherently split Bose gases in the
absence of tunnel coupling. This setup has been extensively studied in
the literature, see e.g. \cite{Kitagawa2010,Kitagawa2011}. The
low-energy limit of this problem is particularly simple, because the
symmetric and antisymmetric sectors decouple, and the relevant
dynamics occurs only in the latter. The Hamiltonian in the
antisymmetric sector is
\begin{align}
{\cal H}_a = \frac{\pi v (\delta \hat{N})^{2}}{2KL}
+ \sum_{q \neq 0} v |q| \hat{a}_{q}^{\dagger} \hat{a}_{q}.
\label{eq:Ham_application}
\end{align}
As we are dealing with a free theory the initial state is fixed by
specifying the two-point function after the splitting process. In
Refs \cite{Kitagawa2010,Kitagawa2011} this was taken to be of the form
\begin{align}
    \left< \frac{\partial_{x} \hat{\theta}(x)}{\pi}\frac{\partial_{y} \hat{\theta}(y)}{\pi} \right>_{a} &= \frac{\rho}{2}\, \delta_{\xi}(x-y), 
\end{align}
where $\delta_{\xi}(x-y)$ is a delta function which is smeared over
the healing length $\xi$. The corresponding state is
\begin{align}
\ket{W}_{a} = \mathcal{N}_{W} \exp \left( \frac{1}{2} \sum_{k \neq 0}
W^{\vphantom{\dagger}}_{k} \hat{a}^{\dagger}_{k}\hat{a}^{\dagger}_{-k} \right) \ket{0}_{a}
\ket{\psi_{k=0}},
\label{eq:init_state}
\end{align}
where
\be
W_{k} = \frac{1-\alpha_{k}}{1+\alpha_{k}}\ ,\qquad
\alpha_k=
\frac{|k|K}{\pi \rho}\ ,\quad \label{eq:alpha_def}
\braket{n | \psi_{k=0}}_{a} = \left( \frac{1}{\pi \rho_{0} L}
\right)^{1/4} \exp \left( - \frac{1}{2 \rho_{0} L} n^{2} \right),
\ee
with $\ket{n}$ the eigenstate of $\delta N_{a}$ with eigenvalue $n$.
To connect as closely as possible to the existing literature
we adopt the choice (\ref{eq:init_state}) in what follows but note
that our analysis can be straightforwardly adapted to other initial
states.

The Hamiltonian in the symmetric sector is of precisely the same form as (\ref{eq:Ham_application}). For simplicity we will assume the symmetric sector to start
out in a Fock state  
\begin{align}
\ket{\psi}_{s}=\ket{\{n_{q}\}}, \label{eq:symm_state}
\end{align}
with occupation numbers that follow a Bose-Einstein distribution
\begin{align}
n_{k} = \frac{1}{e^{\beta v |k|} - 1}\ .
\end{align}
Initializing the symmetric sector in a thermal state is common in the literature \cite{Kitagawa2011} and rests upon the assumption that the symmetric sector is not affected by the splitting procedure, so that it inherits the thermal properties of the gas before splitting. Since (\ref{eq:symm_state}) is an eigenstate of the symmetric sector Hamiltonian, and mixing between sectors does not occur, the symmetric sector will be in the state (\ref{eq:symm_state}) for all times.

In this Section, we will first express the initial state $\ket{W}_{a}$ in terms of the eigenstates $\ket{\{f_{n}\}}_{a}$ of the vertex operator. Using the simple harmonic oscillator form of the Hamiltonian (\ref{eq:Ham_application}), we will then describe time evolution of the overlap coefficients, and interpret these as a probability distribution for the eigenvalues of $\hat{\rho}_{\mathrm{tof}}(x, \vec{r}, t)$, which are directly measured in experiment.

\subsection{Overlap coefficients} 
\label{sub:time_dependent_overlap_coefficients}

\subsubsection{Antisymmetric sector}

The overlap coefficients $\braket{\{f_{n}\}|W(t)}_{a}$ can be represented
as products over the modes. The contributions from the finite momentum
modes are obtained in complete analogy to Appendix
\ref{sec:normalization_of_vertex_operator_eigenstates}. The zero modes
require a separate consideration, which is given in Appendix
\ref{sec:time_dependent_overlap_for_the_zero_mode}. Combining the two
kinds of contributions gives the result 
\begin{align}
\begin{split}
\left|\braket{\{f_{n}\}|W(t_0)}_{a}\right|^{2} = \sqrt{2\pi c_{0}(t_0)} \prod_{k \geq 0} \frac{1}{2\pi c_{k}(t_0)} \exp \left( -\frac{\left( {\rm Re} f_{k} \right)^{2} + \left( {\rm Im} f_{k} \right)^{2} }{2c_{k}(t_0)} \right) ,  
\label{eq:time_dep_overlap}
\end{split}
\end{align}
where we have defined the time-dependent variances
\be
c_k(t_0)=\begin{cases}
\frac{1}{4 \rho_{0} L} \Big(\cos^{2} \left( v|k|t_0 \right) + \left( \frac{k_{c}}{k} \right)^{2} \sin^{2} \left( v|k|t_0 \right)  \Big) & \text{if
}k\neq 0,\\
\frac{1}{2 \rho_{0} L}\Big(1+\left(v k_{c} t_0\right)^2 \Big) &\text{if
}k=0.
\end{cases}
\label{eq:c_variances}
\ee
The momentum scale occurring here is given by $k_{c} = 2 \pi / \xi$, where $\xi$ is the healing length of the gas. Any fluctuations below this length scale are not captured by the low-energy effective Luttinger Liquid theory.

\subsubsection{Symmetric sector}

To describe the effects of longitudinal expansion, operators in the
symmetric sector must be included in the density operator, via
(\ref{eq:density_tof_bos_longitudinal_approx}). In analogy with the
antisymmetric sector, a measurement then corresponds to a projection
to simultaneous eigenstates $\ket{\{f_{q}\}}_{s}$ of
$e^{i\hat{\phi}_{s}(x)}$ in the symmetric sector. These eigenstates
will have the same form as their antisymmetric counterparts, presented
in (\ref{eq:eigenstate}). The probability of measuring the
corresponding eigenvalue $e^{i\varphi_{s}(x)}$ will similarly be given
by the squared overlap with the state of the system in the symmetric
sector. 

Assuming the symmetric sector to occupy the state (\ref{eq:symm_state}) at all times, the overlap coefficients with the eigenstates $\ket{\{f_{q}\}}_{s}$ of $e^{i\hat{\phi}_{s}(x)}$ are computed in Appendix \ref{app:overlap_fock_state}, and read
\begin{align}
\big| \braket{\{f_{q}\}|\psi}_{s} \big|^{2} = \prod_{q>0}\frac{1}{\pi |u_{q}|^{2}} L^{2}_{n_{q}}\left( \Big| \frac{f_{q}}{u_{q}} \Big|^2 \right) e^{- \big| \frac{f_{q}}{u_{q}} \big|^2}, \label{eq:overlap_fock}
\end{align}
where $L_{n}(x)$ is the Laguerre polynomial of degree $n$.

\subsection{Analysis of vertex operator eigenvalue distributions} 
\label{sub:analysis_of_phase_distribution_functions}
\begin{figure}[htbp]
\centering
(a)\includegraphics[width=0.4\textwidth]{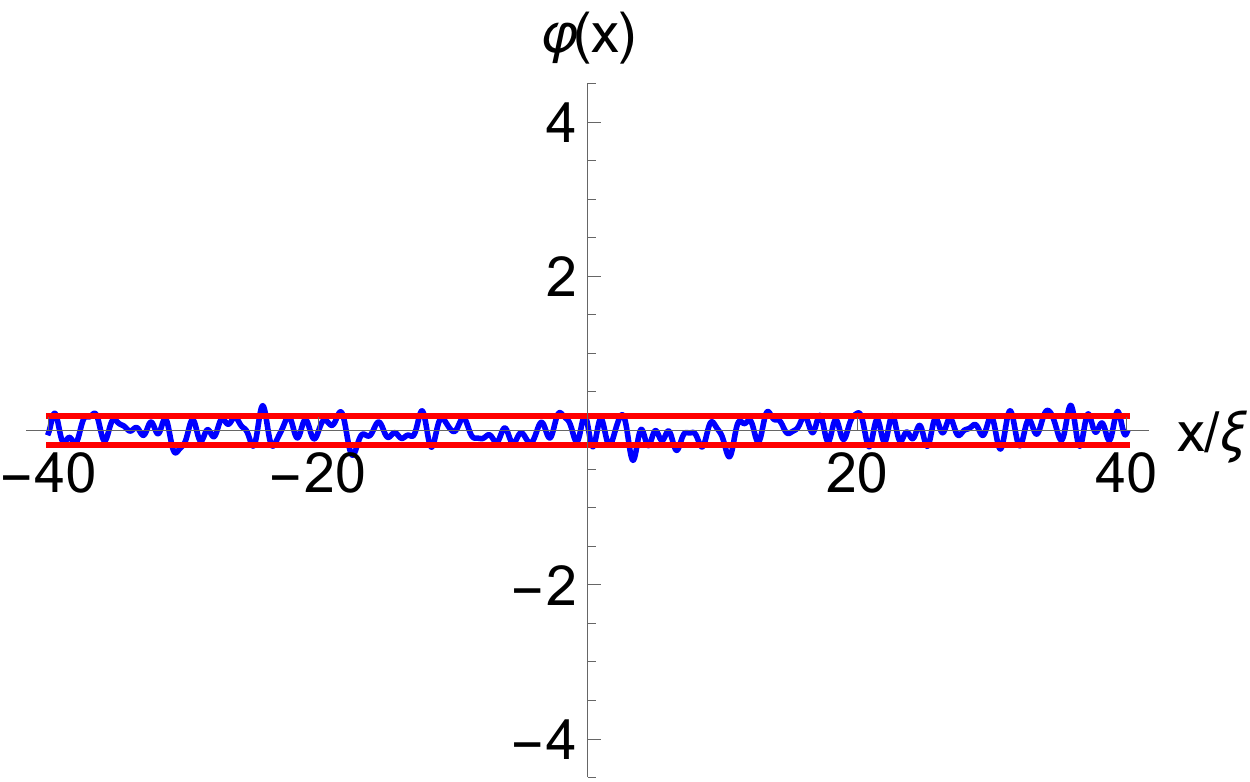}
 \hspace{0.05\textwidth}
(b)\includegraphics[width=0.4\textwidth]{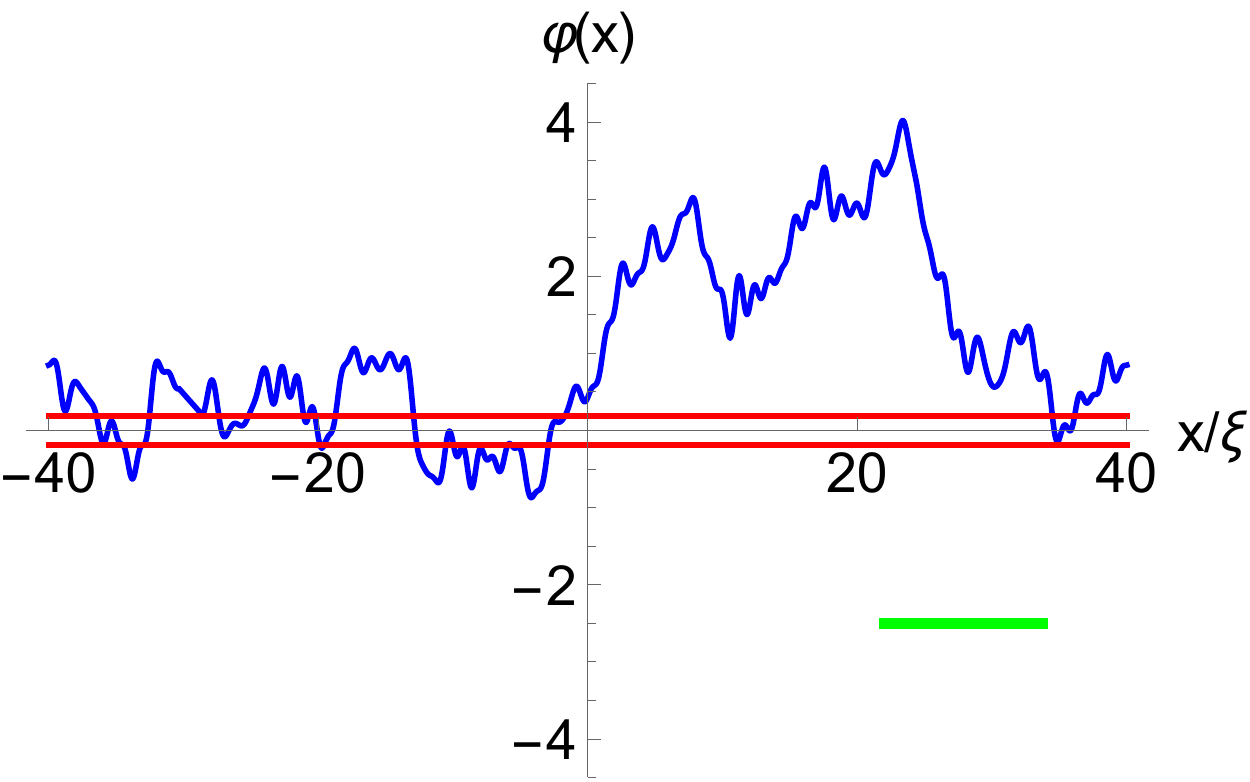}
\caption{Individual realizations of the eigenvalue $\varphi_{a}(x,t_0)$ for
  the phase field $\hat{\phi}_{a}(x,t_0)$. The typical behavior at $t_0=0
  \,\xi/v$, cf. (a), is distinctly different from that at $t_0=14
  \,\xi/v $, cf. (b). At $t_0=0$, small fluctuations occur at all
  lengthscales, with a typical amplitude given by $1/\sqrt{K}$ (red
  lines). At later times, the typical fluctuations are larger for
  longer lengthscales. The crossover length scale from which
  fluctuations become large is indicated with a green bar. In terms of
  Luttinger Liquid parameters, it is predicted \cite{Kitagawa2011} to
  be $l_{0} = 8 K^{2}/\rho \pi^{2}$. A further note about experimental 
  parameters is presented in Section \ref{sub:experimental_parameters}.} 
  \label{fig:phase_realizations} 
\end{figure}

The squared overlap coefficients (\ref{eq:time_dep_overlap}) have a
clear physical interpretation: when measuring
$\hat{\rho}_{\mathrm{tof}}(x, \vec{r}, t_1+t_0)$, the overlap coefficient
$\left|\braket{\{f_{n}\}|W(t_0)}_{a}\right|^{2}$ gives the probability
of collapsing to a state for which $e^{i\hat{\phi}_{a}(x,t_0)}$ has
eigenvalue $e^{i\varphi_{a}(x,t_0)}$, with 
\begin{align}
\varphi_{a}(x,t_0) = \sum_{j} f_{j} \, e^{i p_{j} x} .
\end{align}
Examples of typical configurations $\varphi_{a}(x,t_0)$ are shown in
Fig.~ \ref{fig:phase_realizations}.
We first consider the situation at $t_0=0$. In that case the
coefficients $f_{j}$ are drawn from a Gaussian distribution with mean 
$0$ and variance $c_{0}(t_0) = 1/(4\rho_{0} L)$. This results in a
$\varphi_{a}(x)$ with vanishing average and short-wavelength variations of
size $K^{-1/2}$ as shown in the left panel of
Fig.~\ref{fig:phase_realizations}. For $t>0$ the eigenvalues
$\varphi_{a}(x)$ have the structure shown in right panel of Fig.~
\ref{fig:phase_realizations}. At short wavelengths the variations
remain small, while the long wavelength variations become large. The 
cross-over scale between the two behaviours has been determined by
Kitagawa et al. \cite{Kitagawa2011}, and is given by $l_{0} = 8
K^{2}/\rho \pi^{2}$. It is indicated by a green bar in the right panel
of Fig.~ \ref{fig:phase_realizations}. 

\subsection{Experimental parameters} 
\label{sub:experimental_parameters}

In order to facilitate a comparison with experimental data, we use the following parameters from \cite{Kuhnert2013} in all plots: 
after splitting, each of the two gases has one-dimensional density 
$\rho_{0} = 45 \, \mu \mathrm{m}^{-1}$, healing length $\xi= \hbar\pi / mv = \pi \times
0.42 \,\mu \mathrm{m}$ and longitudinal size $L = 80 \, \xi$. When applied to Rubidium atoms, 
this translates to $L \approx 106 \, \mathrm{\mu m}$, with a sound velocity given by $v \approx 1.738 \cdot 10^{-3} \, \mathrm{m}/\mathrm{s}$.
The symmetric sector is in a thermal state, for which we choose $k_{\mathrm{B}}T$ to be some fraction of $\hbar \omega_{\perp}$, with transverse trapping
frequency $\hbar \omega_{\perp} = 2 \pi \times 1.4 \, \mathrm{kHz}$. The state (\ref{eq:init_state}) of the antisymmetric sector is not thermal, but it has an energy density given by $\pi v/(3\xi^{2})$. To compare this to the energy scale of the symmetric sector, we note that a thermal state with the same energy would be at a temperature of approximately $14 \, \mathrm{nK}$, for the parameters presented here.


\section{Results for density measurements after expansion} 
\label{sub:results}
We now return to the (approximate) expression for the gas density
after time of flight, given by (\ref{rhoTOFlargeK}), 
\begin{align}
\begin{split}
\hat{\rho}_{\mathrm{tof}}(x,\vec{r},t_{1}+t_0)\cong 2 \rho_{0} |A_0|^2
\left|f( \vec{r},t_{1})\right|^{2} \left( 1 + \mathrm{Re} \left[
  e^{i\hat{\phi}_{a}(x,t_0)+ i\vec{d}\cdot \vec{r}\, m/t_{1}} \right]
\right). \label{eq:density_tof_bos_final} 
\end{split}
\end{align}
which is valid when longitudinal expansion can be neglected (in the
general case one instead uses (\ref{eq:density_tof_bos_longitudinal_approx})).
A measurement causes the system to collapse to an eigenstate of this
operator and concomitantly a simultaneous eigenstate of
$e^{i\hat{\phi}_{a}(x,t_0)}$. The measurement outcome corresponds to the eigenvalues
\begin{align}
\begin{split}
\varrho_{\mathrm{tof}}(x,\vec{r},t_{1}+t_0)\cong 2 \rho_{0} |A_0|^2 \left|f( \vec{r},t_{1})\right|^{2} \left( 1 + \mathrm{Re} \left[ e^{i\varphi_{a}(x,t_0)+ i\vec{d}\cdot \vec{r}\, m/t_{1}} \right] \right),
\end{split}
\label{eigenvalue_simple}
\end{align}
where $\varphi_{a}(x,t_0)$ is characterized by its Fourier coefficients
$f_{k}$. The probability to measure an eigenvalue
$\rho_{\mathrm{tof}}(\vec{r},t_{1}+t_0)$ with a corresponding set of Fourier
coefficients $\{f_{k}\}$ is given by the overlap coefficient with the
state of the system at the time of release. These overlap coefficients
can be computed in specific cases, as we have demonstrated for the
case of coherently split bose gases without tunnel-coupling, presented
in (\ref{eq:time_dep_overlap}). A completely analogous procedure
can be used to describe a measurement of the observable in eqn
(\ref{eq:density_tof_bos_longitudinal_approx}), which requires
additional overlaps in the symmetric sector, such as those presented
in (\ref{eq:overlap_fock}). 

\begin{figure}[htbp]
\centering
\includegraphics[width=0.32\textwidth]{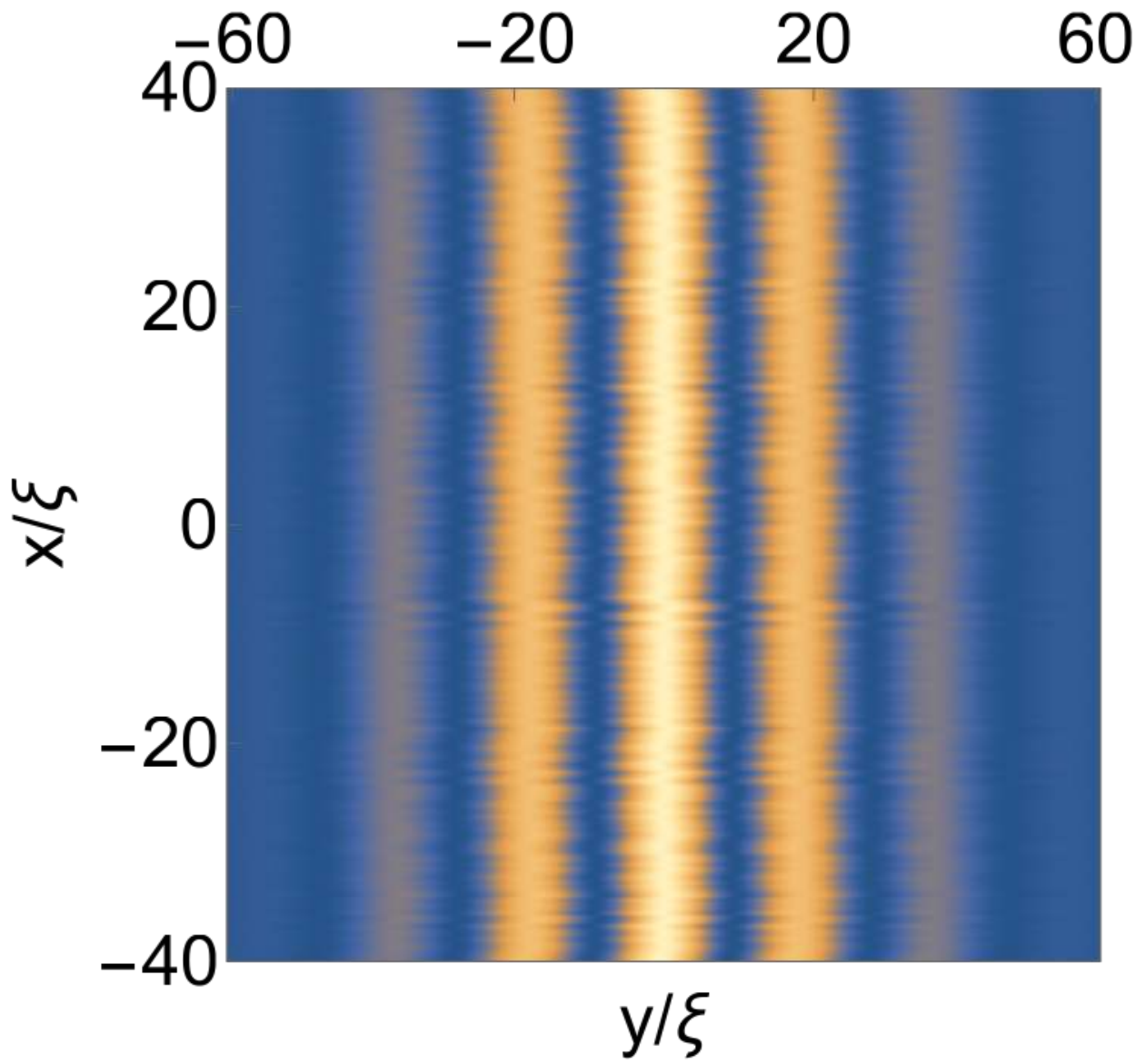}
\includegraphics[width=0.32\textwidth]{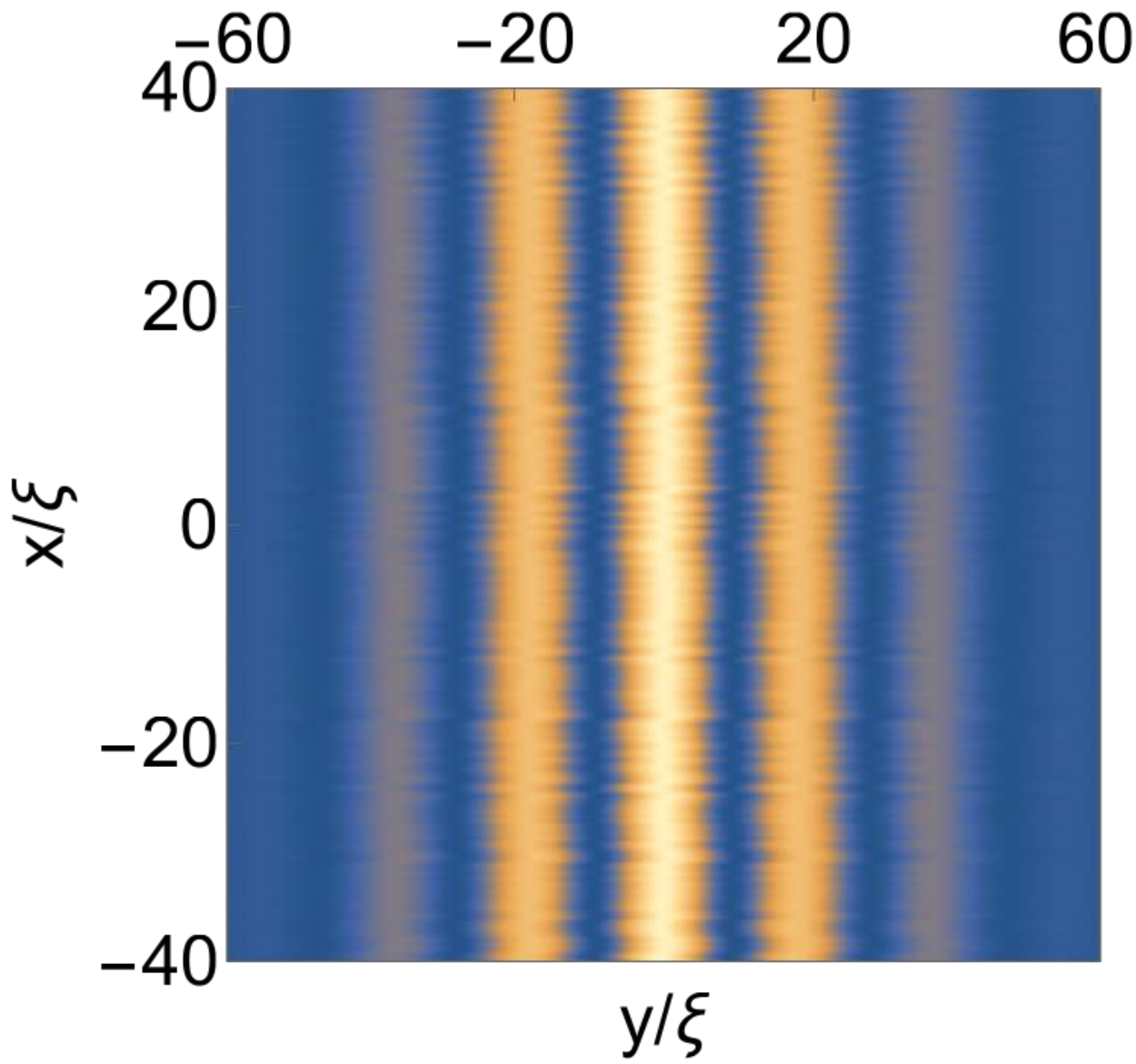}
\includegraphics[width=0.32\textwidth]{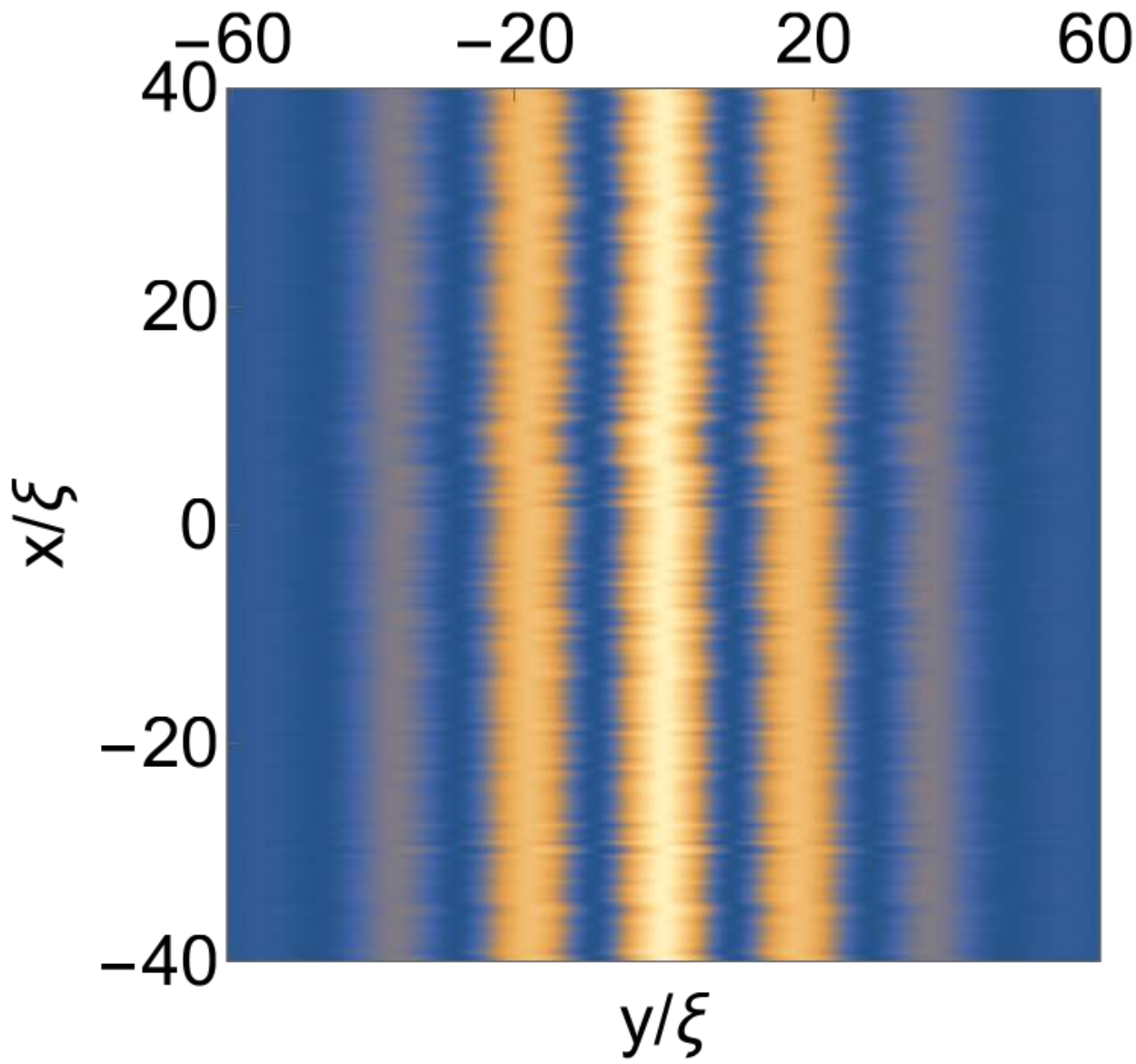}
\caption{Samples of outcomes for individual (simultaneous) measurements of
  $\hat{\rho}_{\mathrm{tof}}(\vec{r},t_{1}+t_0)$ at $t_0=0$, using (\ref{eq:density_tof_bos_final}). The parameters
are as presented in Section \ref{sub:experimental_parameters} and
the time of flight is taken as $t_1=16 \, \mathrm{ms}$.}  
  \label{fig:Dens}
\end{figure}

\begin{figure}[htbp]
\centering
\includegraphics[width=0.32\textwidth]{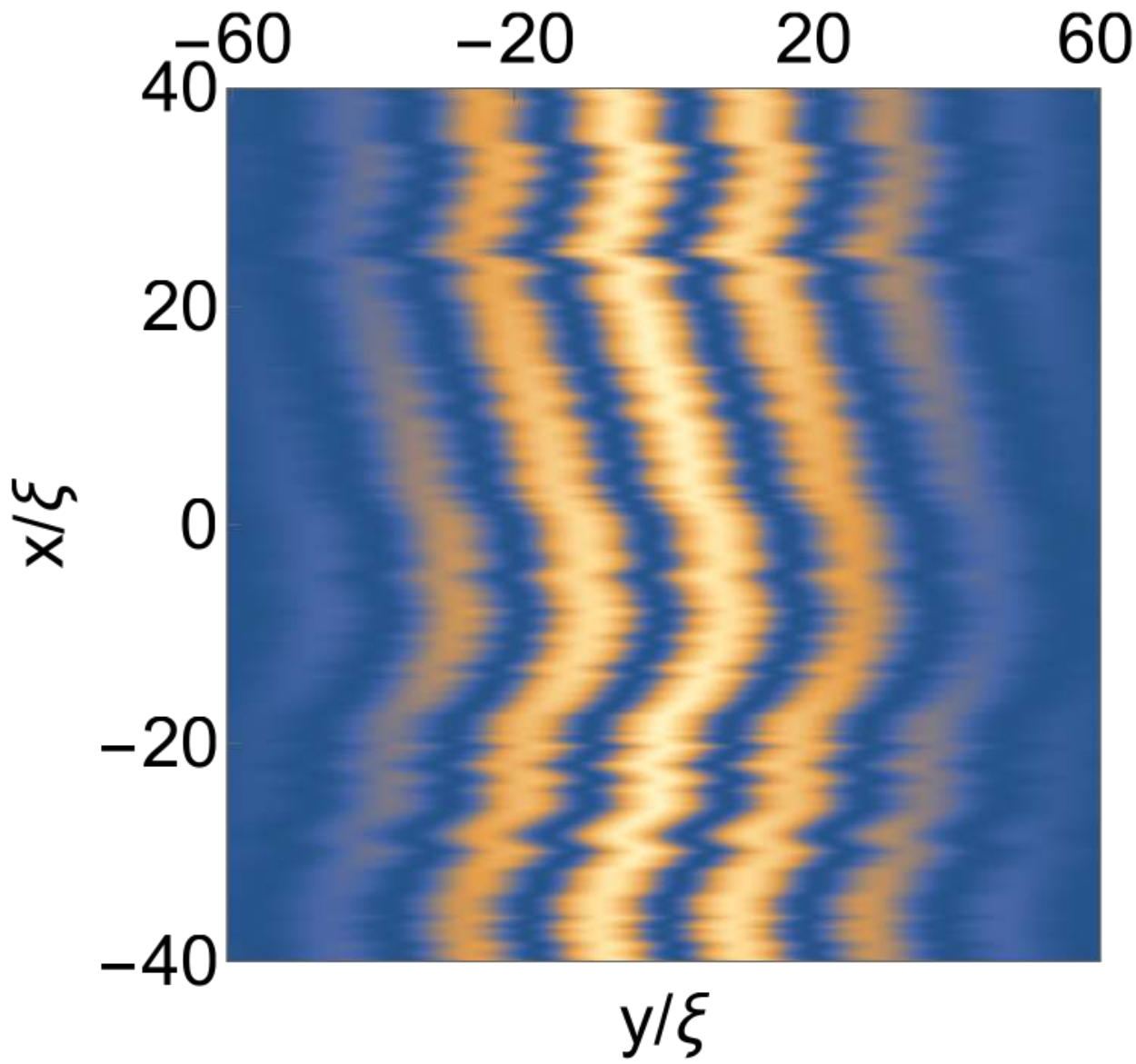}
\includegraphics[width=0.32\textwidth]{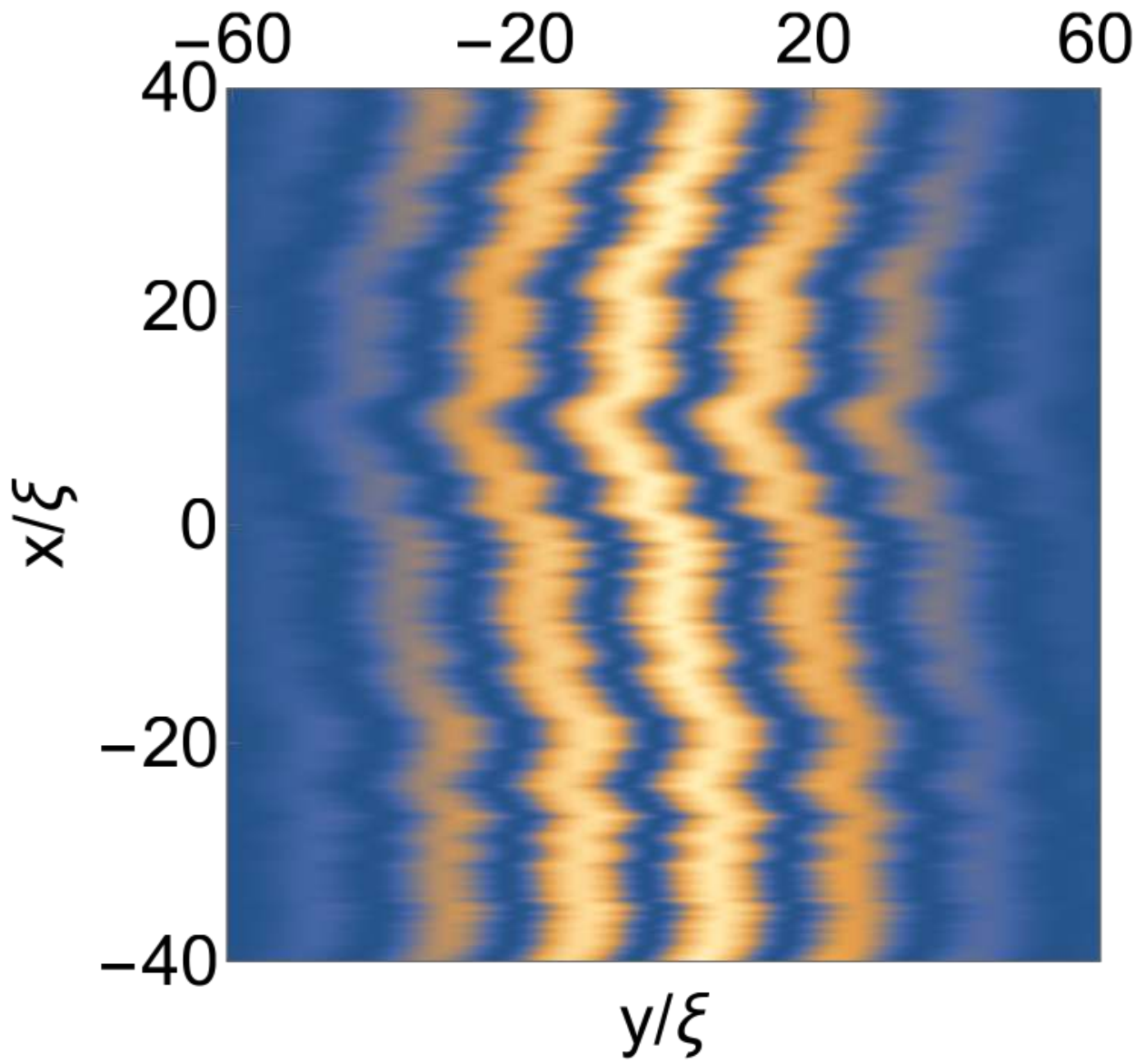}
\includegraphics[width=0.32\textwidth]{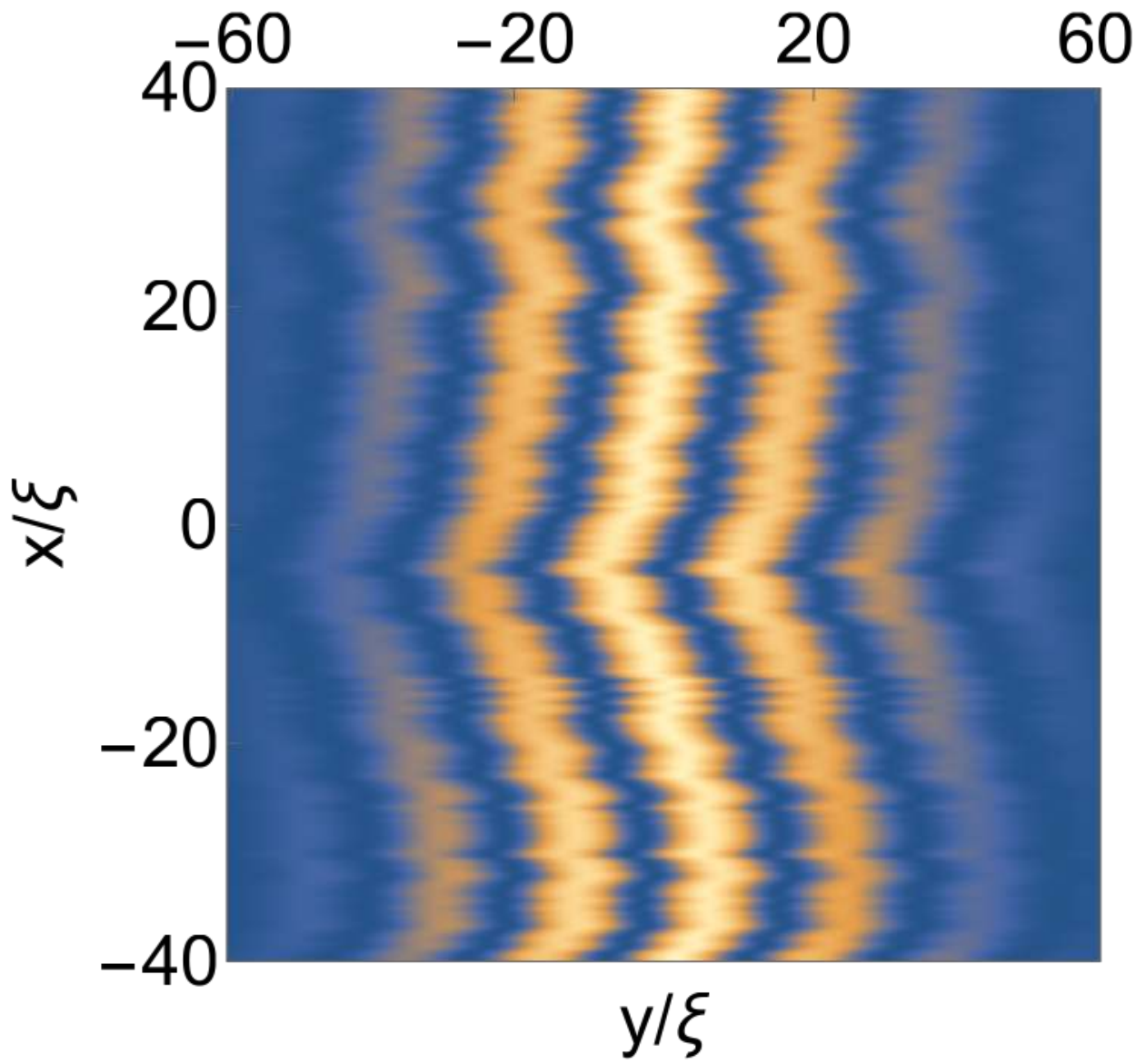}
\caption{Samples of outcomes for individual (simultaneous) measurements of
  $\hat{\rho}_{\mathrm{tof}}(\vec{r},t_{1}+t_0)$ at $t_0=14
  \,\xi/v \approx 10.6 \, \mathrm{ms}$, using (\ref{eq:density_tof_bos_final}). The parameters
are as presented in Section \ref{sub:experimental_parameters} and the time of flight is taken as
$t_1=16 \, \mathrm{ms}$.}   
\label{fig:Dens2}
\end{figure}

With the above formalism in place, experiments can then be modelled
as follows. We assume that our system is initialized in the state
\be
|\Psi(0)\rangle=|W\rangle_a\otimes|\psi\rangle_s\ , \label{eq:state_tensor_prod}
\ee
where $|W\rangle_a$ and $|\psi\rangle_s$ are given in
(\ref{eq:init_state}) and (\ref{eq:symm_state}) respectively. We then
let the system evolve under the Luttinger liquid Hamiltonian
\fr{eq:Ham_Bosonized} for a time $t_0$. At time $t_0$ we switch the
time evolution to a free expansion and perform a projective density
measurement at time $t_0+t_1$.
Some representative results for
$\rho_{\mathrm{tof}}(x,\vec{r},t_{1}+t_0)$ evaluated using the
simplified expression \fr{eigenvalue_simple} are presented  
in Figs~\ref{fig:Dens} and \ref{fig:Dens2}. Here the time of flight is
taken to be $t_1=16\, \mathrm{ms}$. 

\begin{figure}[htbp]
\centering
\includegraphics[width=0.32\textwidth]{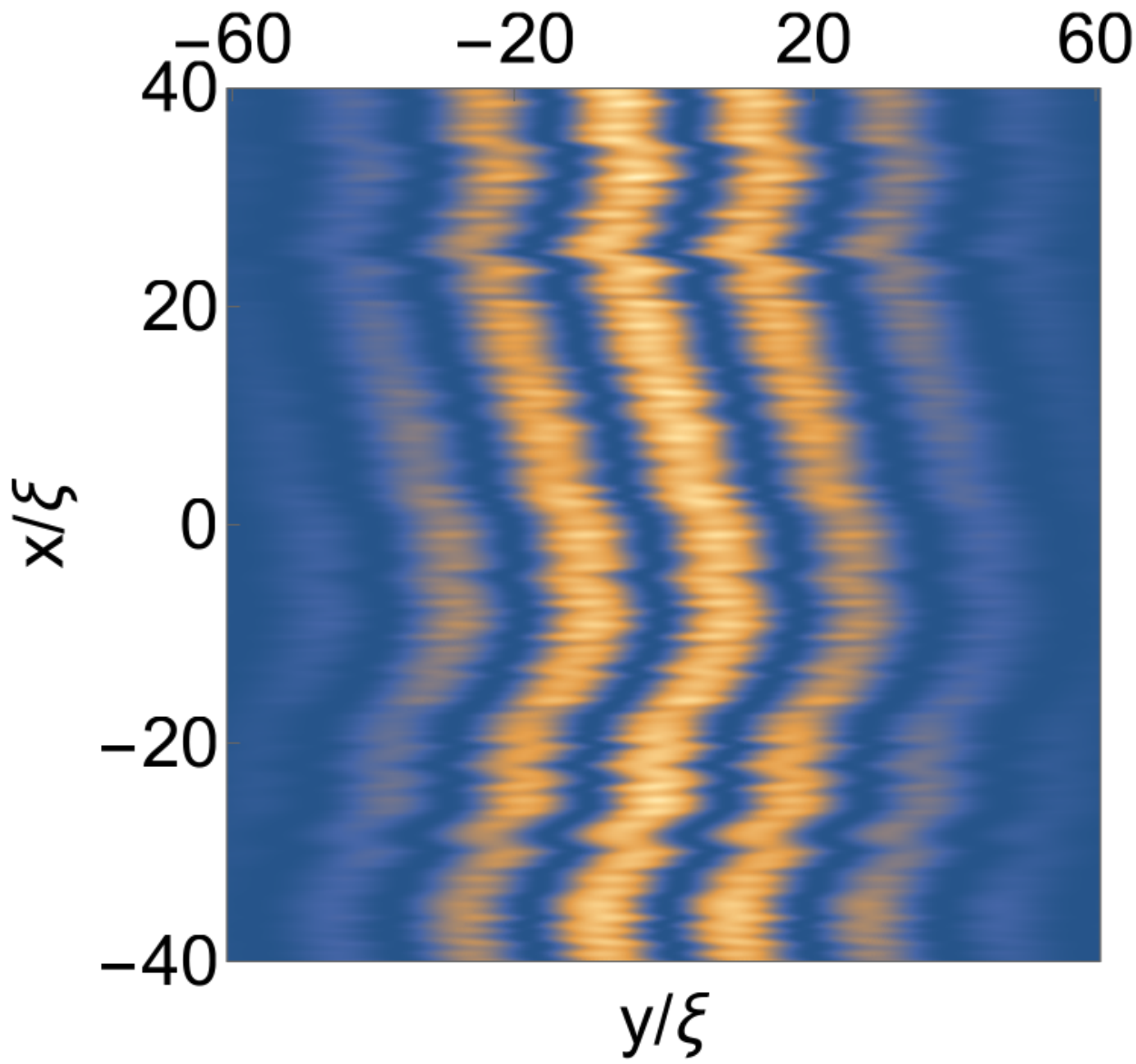}
\includegraphics[width=0.32\textwidth]{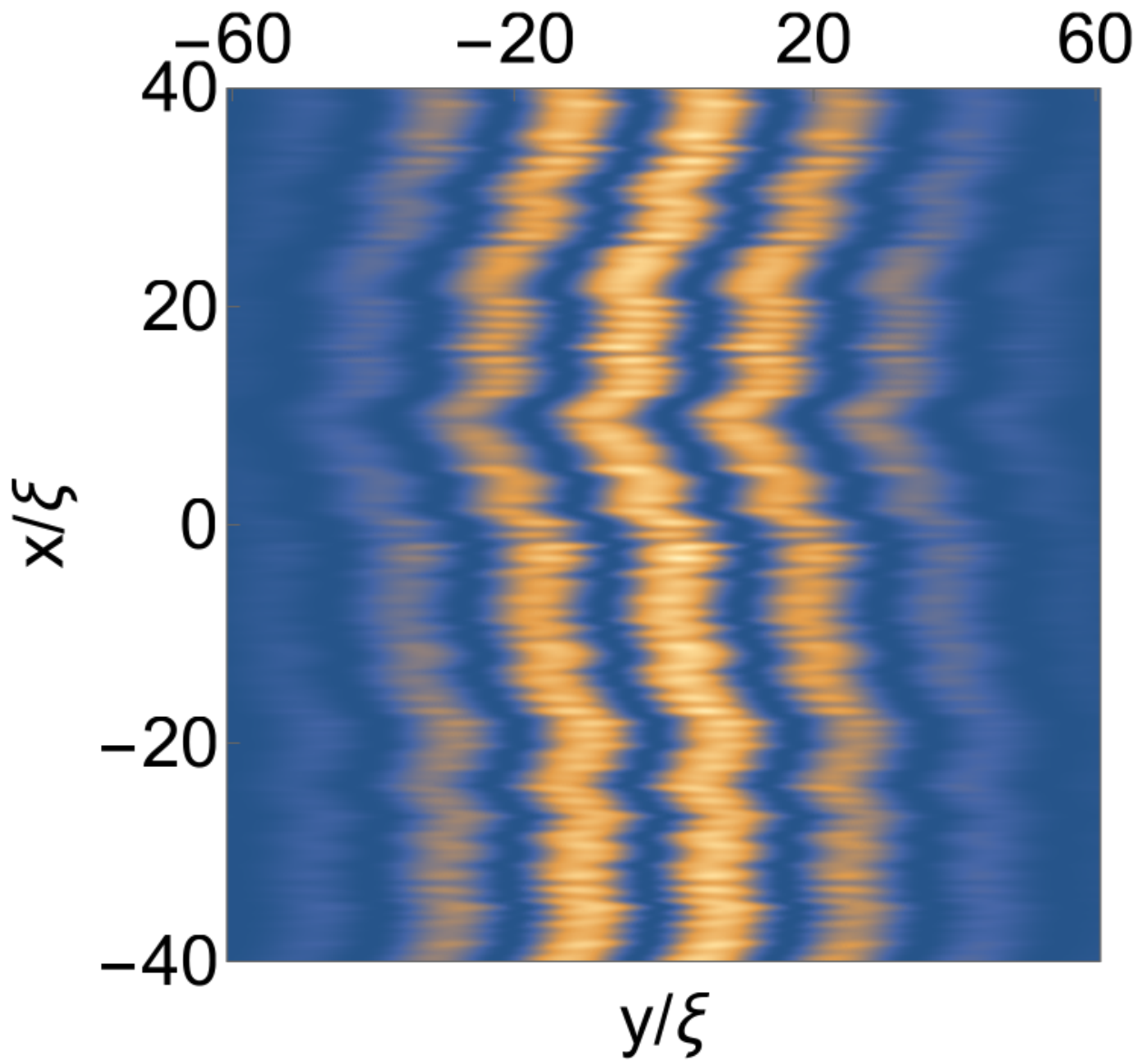}
\includegraphics[width=0.32\textwidth]{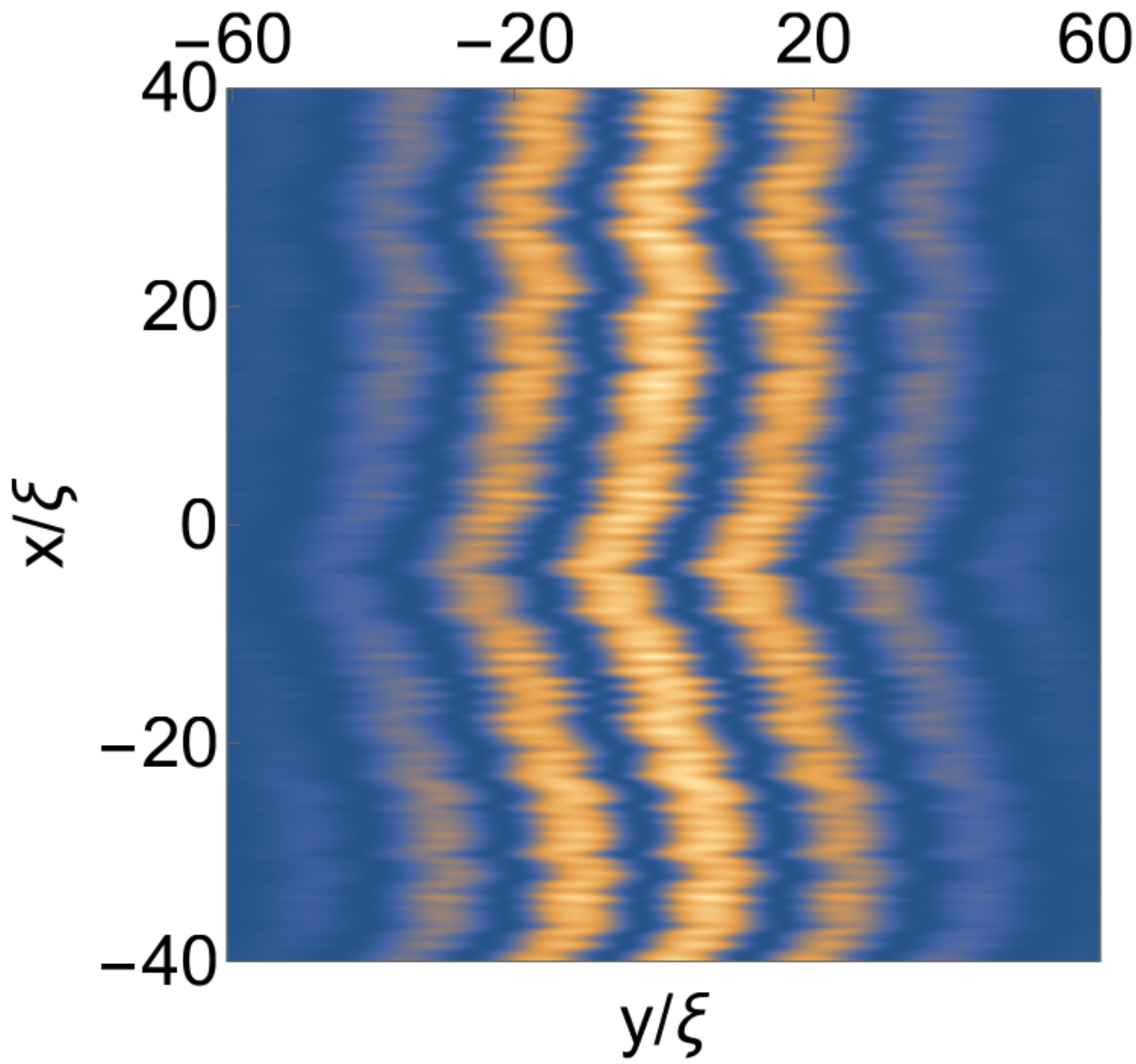}
\caption{Samples of outcomes for individual (simultaneous) measurements of
  $\hat{\rho}_{\mathrm{tof}}(\vec{r},t_{1}+t_0)$ at $t_0=14
  \,\xi/v$ and $t_{1} = 16 \, \mathrm{ms}$. The plots were produced using
  \fr{rhoTOFlargeK}, including the term proportional to 
$\partial_x\hat\theta_s$. The temperature in the symmetric sector is
$34 \, \mathrm{nK}$, which corresponds to
$k_{\mathrm{B}}T = 0.5 \, \hbar \omega_{\perp}$, using the parameters
 presented in Section \ref{sub:experimental_parameters}.}  
  \label{fig:Dens_3shots_full}
\end{figure}

We see that after a sufficiently long time of flight the measured
density exhibits a number of ``interference fringes'' in the
transverse direction. In the initial state ($t_0=0$) these are
straight, but if the split condensate is left to time evolve ($t_0>0$)
they start bending. We stress that the intensity along a given fringe
does not vary with $x$. This is a property of the simplified
expression \fr{eigenvalue_simple} which assumes that the longitudinal
expansion and the density fluctuations in the symmetric sector are
negligible. Retaining the term proportional to
$\partial_x\hat\theta_s$ in \fr{rhoTOFlargeK} does introduce
variations in the intensity of the individual fringes. Examples of
such realizations are presented in Fig.~\ref{fig:Dens_3shots_full}. 

\subsection{Effects of the longitudinal expansion}

When the effects of longitudinal expansion are included via
(\ref{eq:density_tof_bos_longitudinal_approx}) the measured density
operator is no longer exclusively a function of the relative phase
operator but now includes the phase operator from the symmetric sector
as well. This dependence on $e^{i\hat{\phi}_{s}(x)}$ is modeled in
complete analogy to our discussion of $e^{i\hat{\phi}_{a}(x)}$: we
construct its eigenstates, compute their squared overlap with the
state of the system (\ref{eq:state_tensor_prod}), and interpret this as
a probability distribution for the corresponding eigenvalues. 

\begin{figure}[htbp]
\centering
\includegraphics[width=0.32\textwidth]{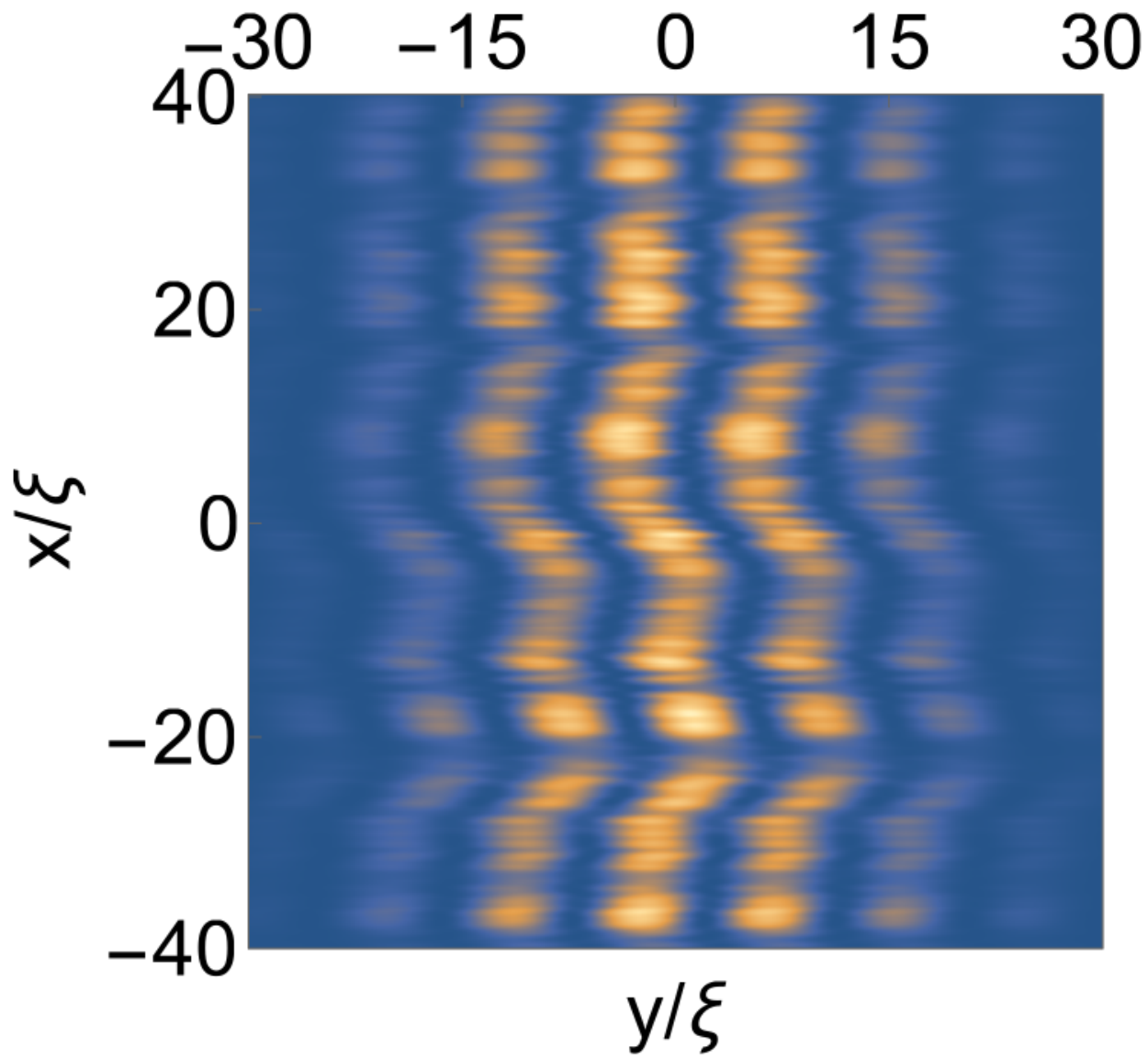}
\includegraphics[width=0.32\textwidth]{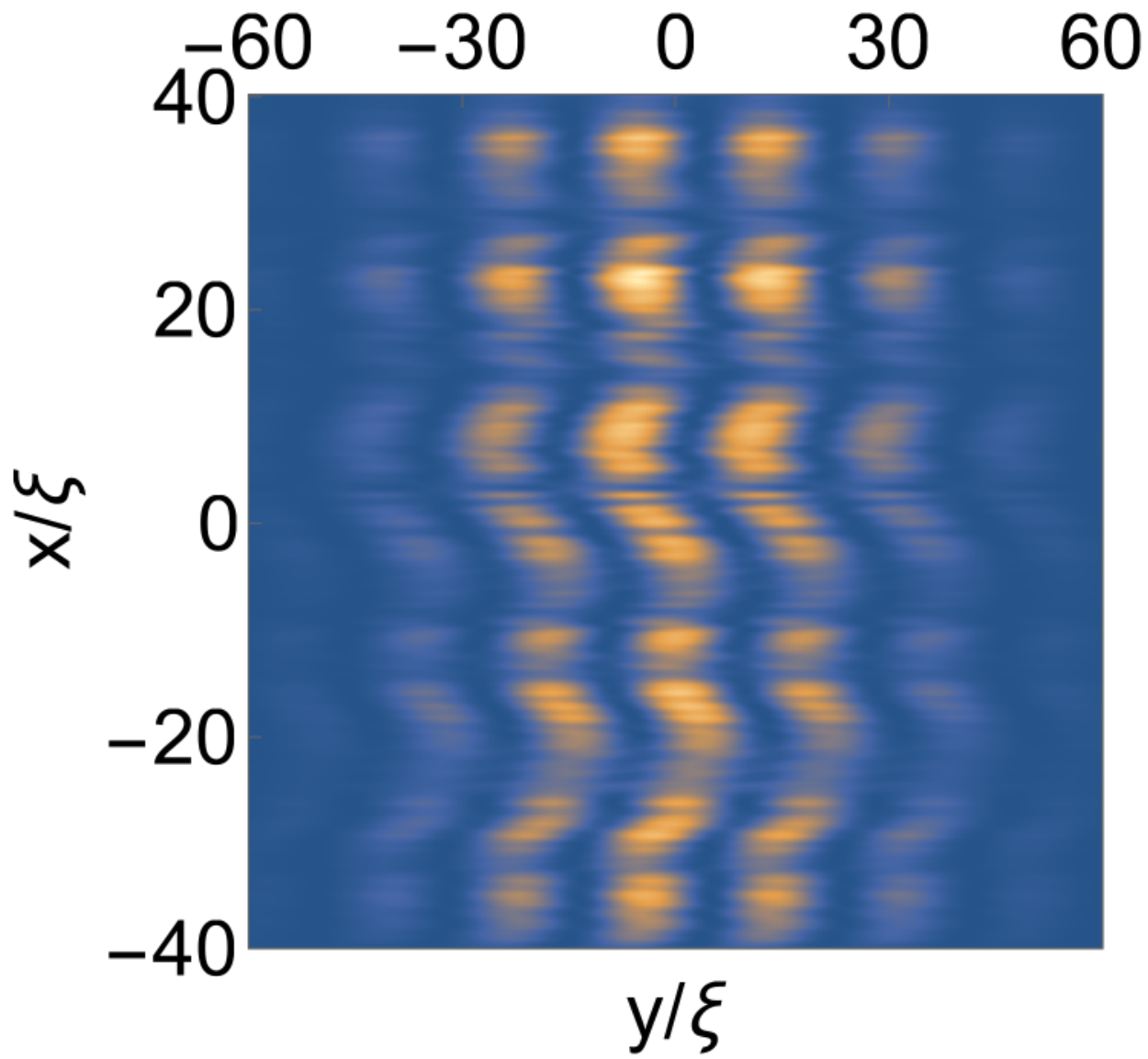}
\includegraphics[width=0.31\textwidth]{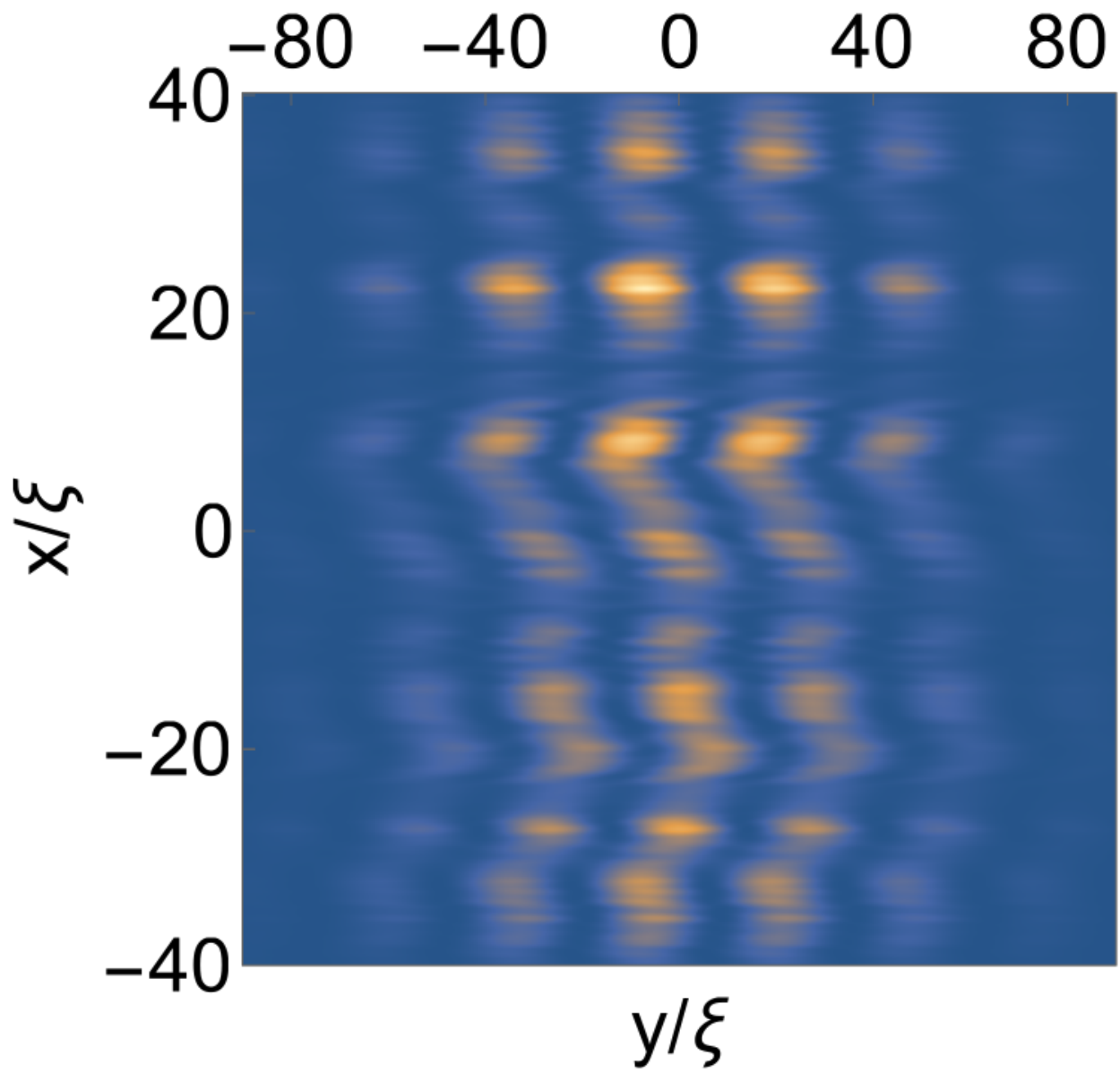}
\caption{Outcomes for a single projective measurement of
$\hat{\rho}_{\mathrm{tof}}(\vec{r},t_{1}+t_0)$, using (\ref{eq:density_tof_bos_longitudinal_approx}), observed for
different time-of-flight values $t_{1}$ and at fixed one-dimensional evolution time $t_{0}= 14 \,\xi/v$. The temperature in the symmetric sector is
$34 \, \mathrm{nK}$, which corresponds to
$k_{\mathrm{B}}T = 0.5 \, \hbar \omega_{\perp}$, using the parameters
 from Section \ref{sub:experimental_parameters}. From left to right, the time of flight is $t_{1} =
8, 16$ and $24 \, \mathrm{ms},$ respectively. The underlying
eigenvalues $e^{i\varphi_{a,s}(x,t_0)}$ are taken to be identical in
all three plots in order to accentuate the effects of the time of flight.}   
\label{fig:Dens_3shots_full2}
\end{figure}

\begin{figure}[htbp]
\centering
\includegraphics[width=0.32\textwidth]{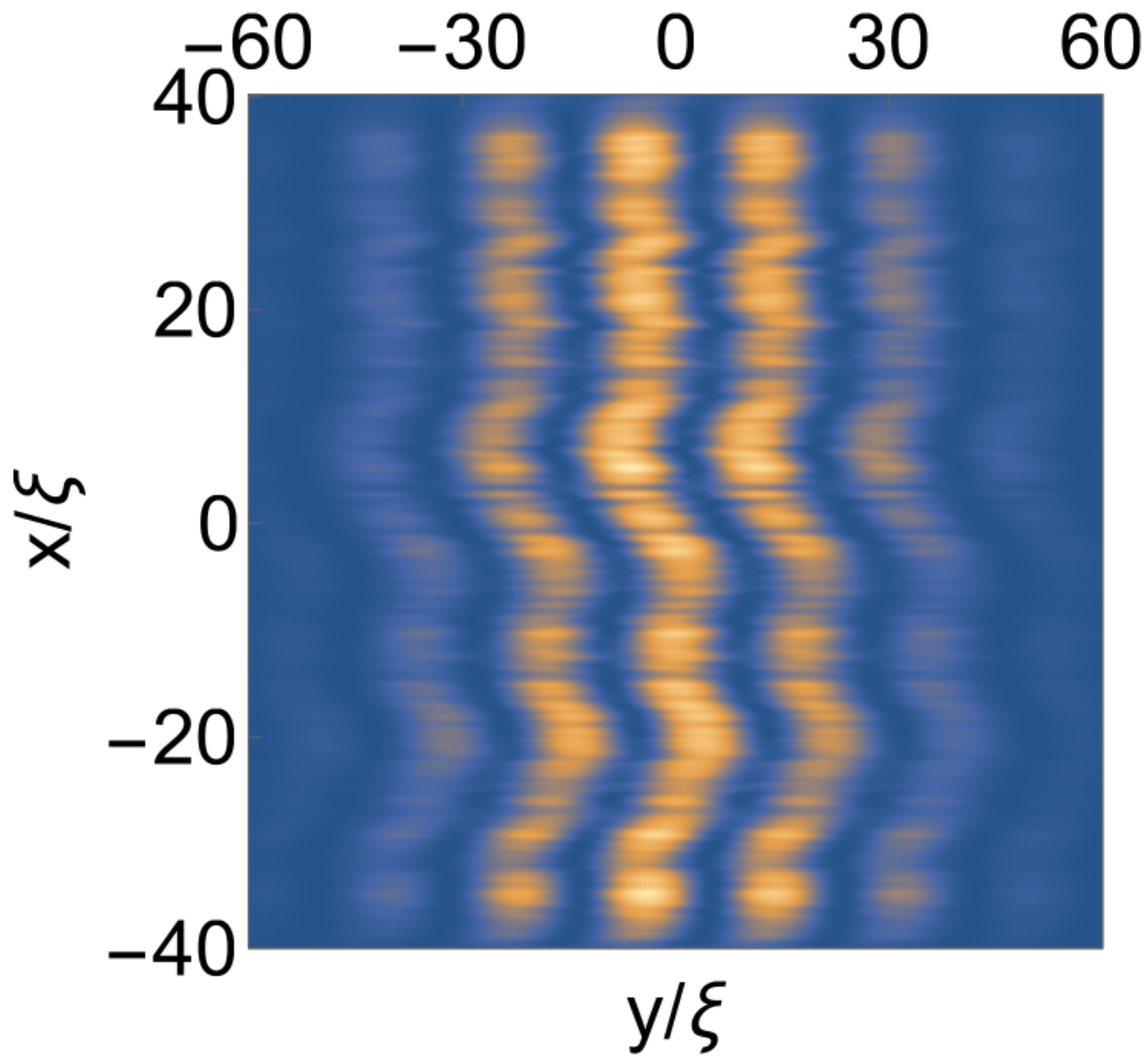}
\includegraphics[width=0.32\textwidth]{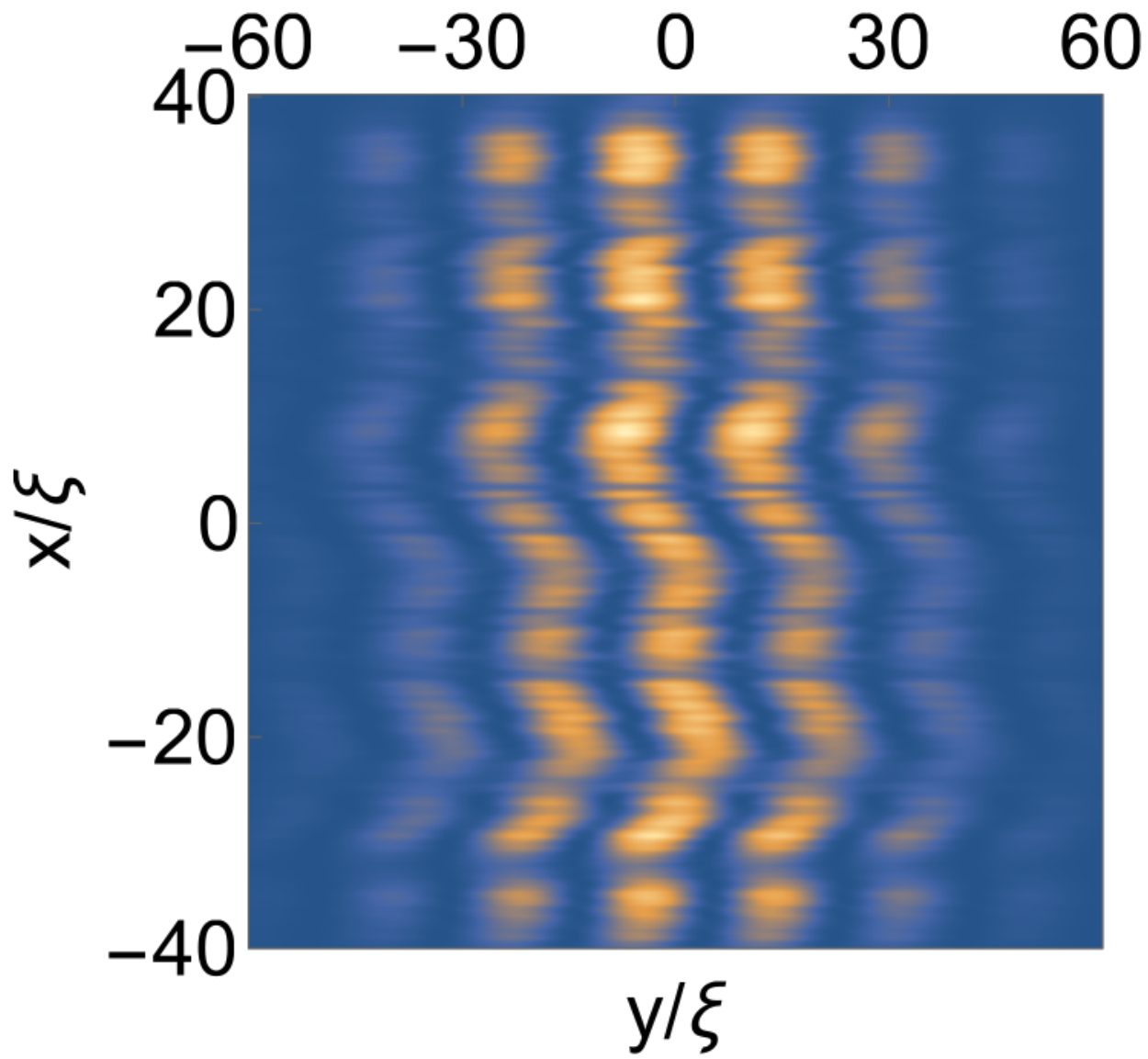}
\includegraphics[width=0.32\textwidth]{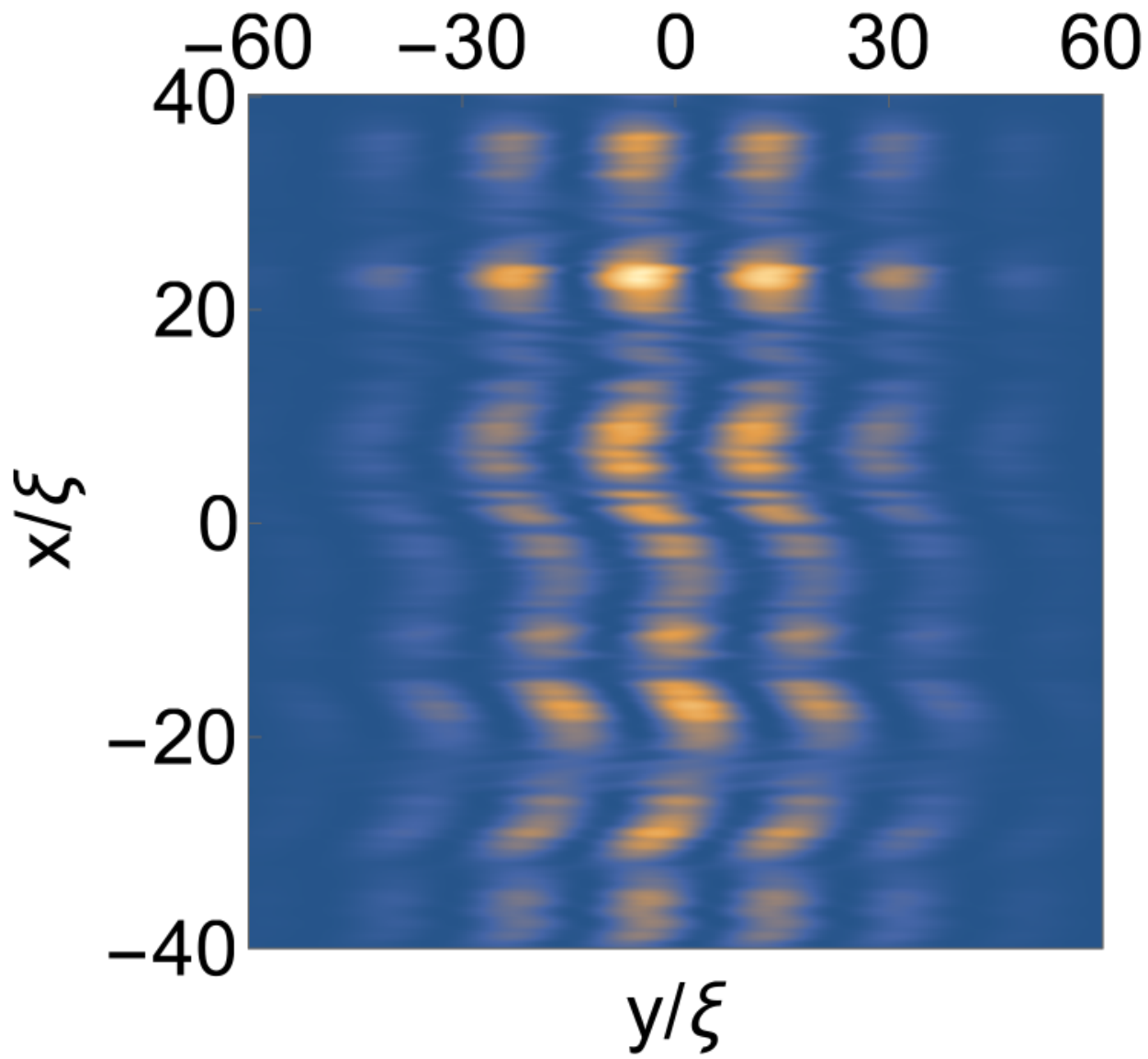}
\caption{Outcomes for projective measurements of $\hat{\rho}_{\mathrm{tof}}(\vec{r},t_{1}+t_0)$, with $t_{1} = 16 \, \mathrm{ms}$ and $t_{0}= 14 \,\xi/v$, created using (\ref{eq:density_tof_bos_longitudinal_approx}). From left to right, the temperatures are $k_{\mathrm{B}}T = 0.1 \, \hbar \omega_{\perp}$, $k_{\mathrm{B}}T = 0.3 \, \hbar \omega_{\perp}$ and $k_{\mathrm{B}}T = 0.7 \, \hbar \omega_{\perp}$. To allow for an easy comparison, the same eigenvalue $\varphi_{a}(x)$ has been used throughout, whereas the eigenvalues $\varphi_{s}(x)$ are drawn from shot to shot, using the temperature-dependent distribution functions for the symmetric sector computed in Section \ref{sub:time_dependent_overlap_coefficients}. The other parameters used here are as presented in Section \ref{sub:experimental_parameters}.}  
\label{fig:Dens_3shots_full2_temps}
\end{figure}

A comparison between this improved analysis (which employs the
overlaps computed in Section
\ref{sub:time_dependent_overlap_coefficients}) and the case of frozen
longitudinal dynamics is presented in Fig.~\ref{fig:Dens_3shots_full2}. It 
can be observed that additional ``density ripples'' emerge in the
longitudinal direction, as a consequence of interference between
points with different longitudinal coordinates in the original two
gases. These density ripples become more pronounced as the time of
flight $t_{1}$ increases, and they occur on longer length scales:
whereas $\varrho_{\rm tof}(x,\vec{r},t_0+t_1)$ only involves operators
at position $x$ at $t_{1}=0$, it acquires contributions from points at
an increasingly large longitudinal separation as $t_{1}$ increases. 
This effect is sensitive to the temperature in the symmetric sector,
as is illustrated in Fig. \ref{fig:Dens_3shots_full2_temps}. 
A detailed analysis of these density ripples in the density-density
correlation function, including their temperature dependence,
has been presented in \cite{Imambekov2009,Manz2010}.

\subsubsection{\texorpdfstring{On extracting the eigenvalues $e^{i\varphi_{a}(x,t_0)}$ from
${\varrho}_{\mathrm{tof}}(\vec{r},t_{1}+t_0)$}{On extracting the eigenvalues of the phase vertex operator from the measured density profile}}

\begin{figure}[htbp]
\centering
(a)\includegraphics[width=0.4\textwidth]{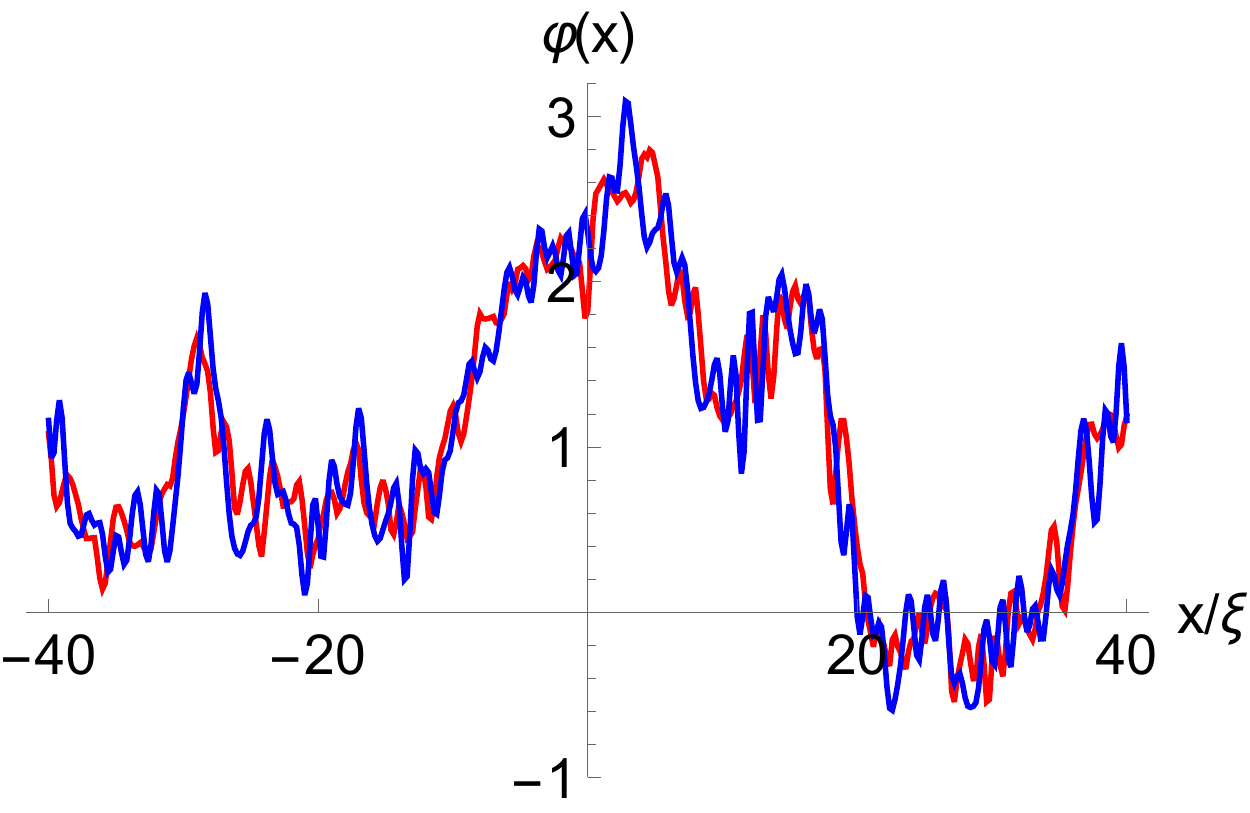}
\hspace{0.05\textwidth}
(b)\includegraphics[width=0.4\textwidth]{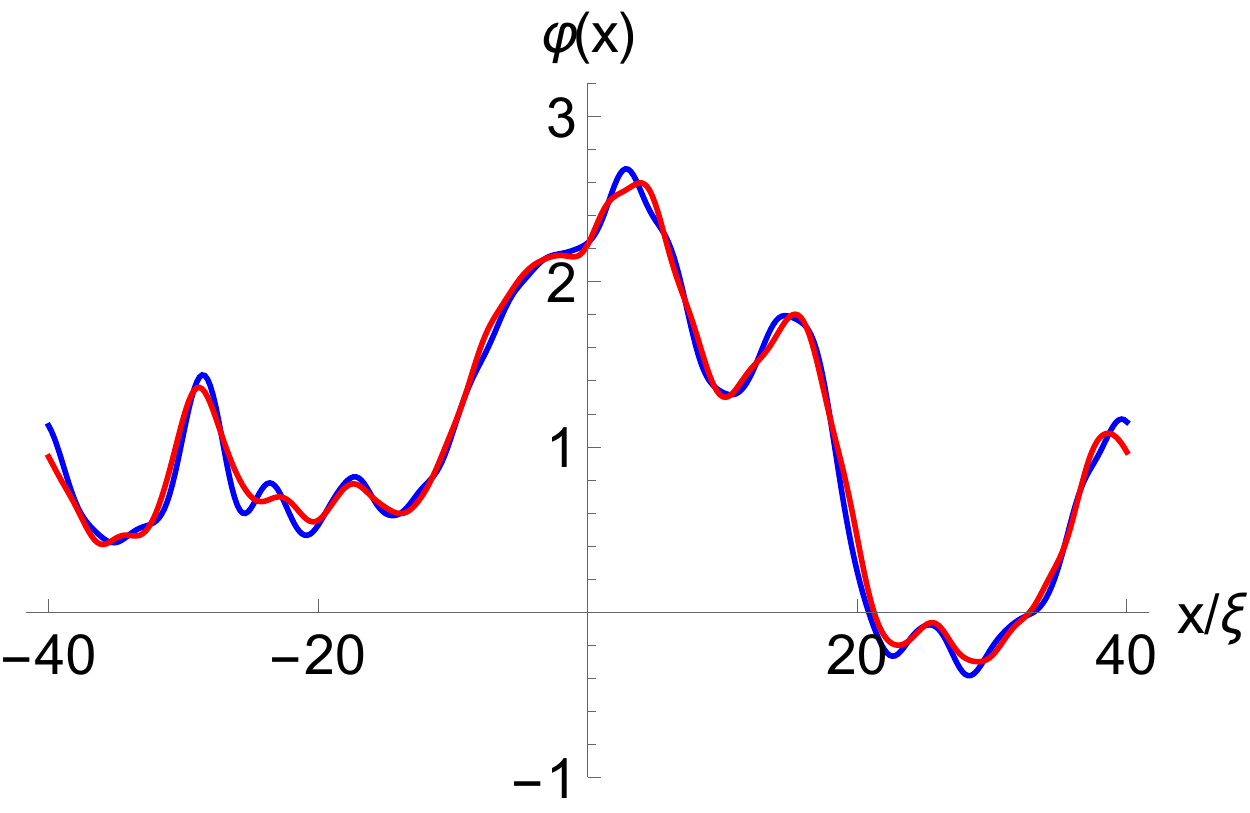}
\caption{(a) Individual realization of the
eigenvalue $\varphi_{a}(x)$ (\textit{blue}), compared to the
extracted phase $\tilde{\varphi}_{a}(x)$ (\textit{red}) at time of
flight $t_{1} = 4 \, \mathrm{ms}$ and $t_{0} = 14 \, \xi/v$.
(b) The same objects, convolved with a Gaussian kernel of width
$\xi = \hbar \pi / m v$. The parameters used here are presented in 
Section \ref{sub:experimental_parameters}, with
$k_{\mathrm{B}}T = 0.5 \, \hbar \omega_{\perp}$, so that $T \approx 34 \,\mathrm{nK}$.} 
\label{fig:extracted_phases_early}
\end{figure}

\begin{figure}[htbp]
  \centering
  (a)\includegraphics[width=0.4\textwidth]{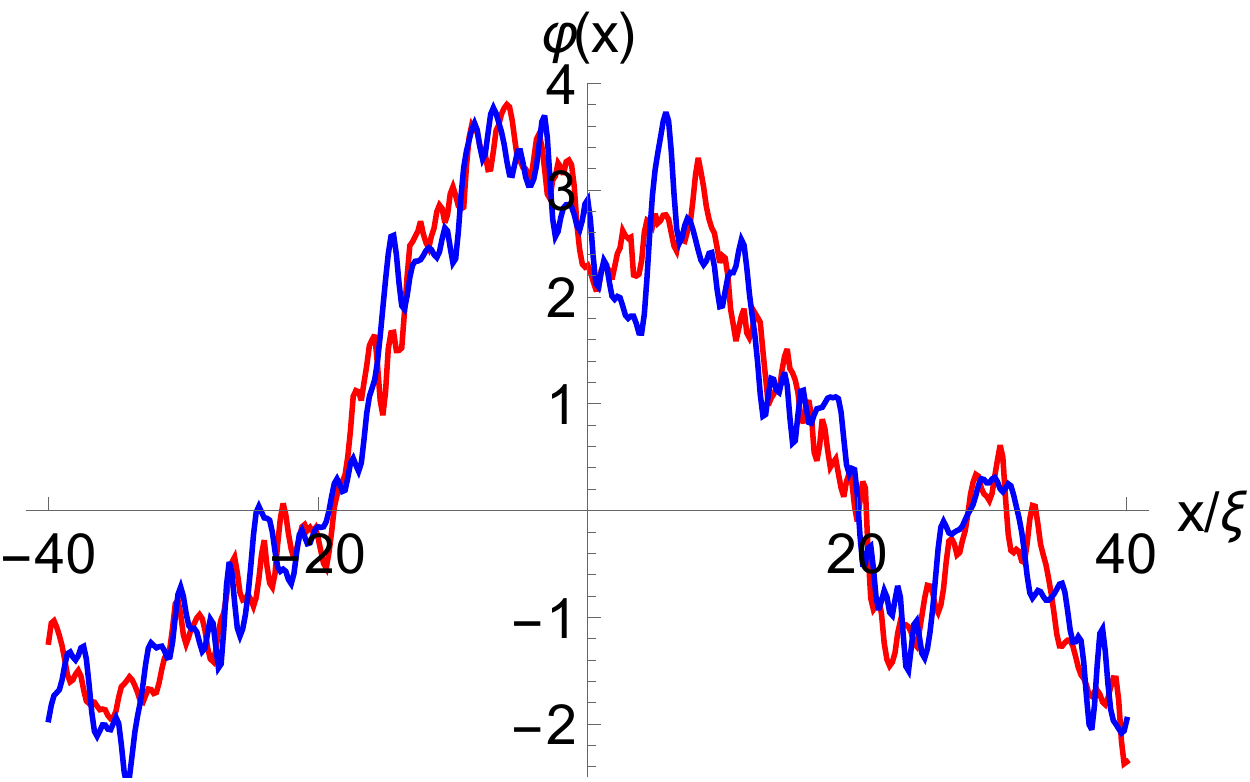}
  \hspace{0.05\textwidth}
  (b)\includegraphics[width=0.4\textwidth]{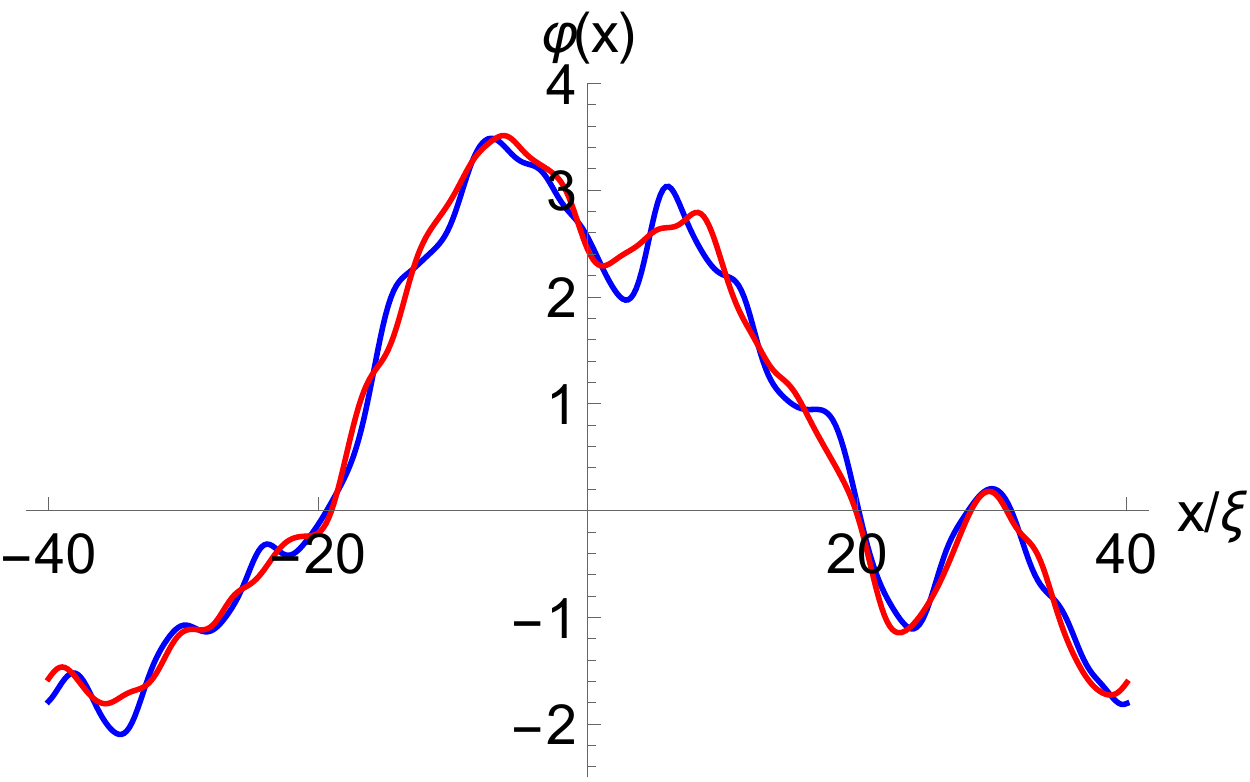}
  \caption{The same as Fig.~\ref{fig:extracted_phases_early}, but at time of flight $t_{1} = 8 \, \mathrm{ms}$, and for a different phase eigenvalue.}
  \label{fig:extracted_phases_medium}
\end{figure}

\begin{figure}[htbp]
  \centering
  (a)\includegraphics[width=0.4\textwidth]{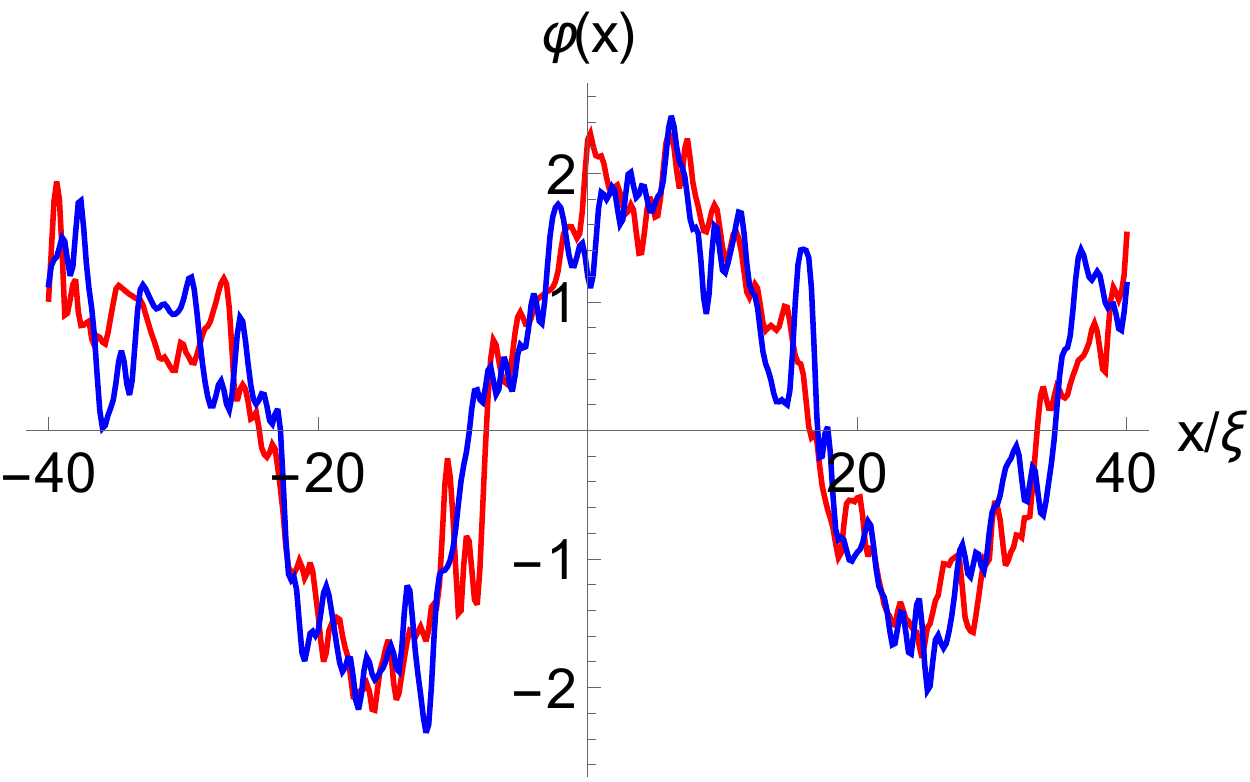}
  \hspace{0.05\textwidth}
  (b)\includegraphics[width=0.4\textwidth]{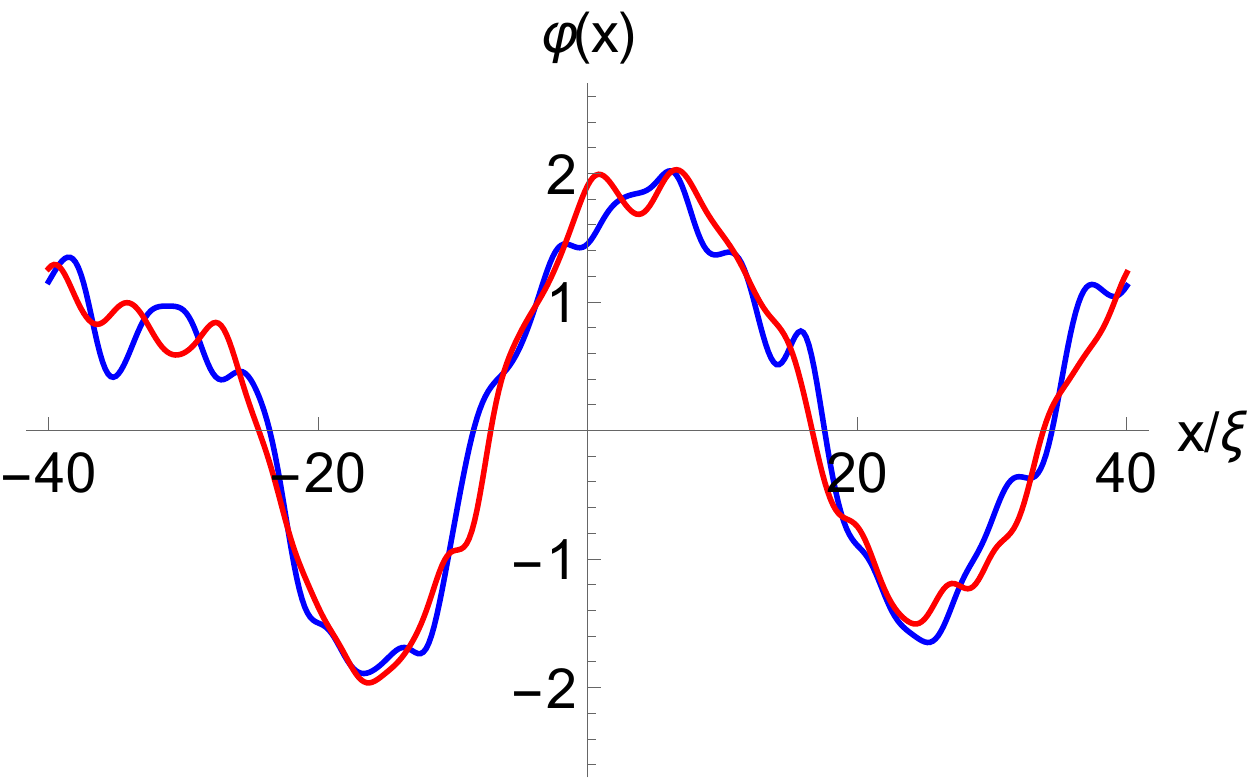}
  \caption{The same as Fig.~\ref{fig:extracted_phases_early}, but at time of flight $t_{1} = 16 \, \mathrm{ms}$, and for a different phase eigenvalue.}
  \label{fig:extracted_phases_medium2}
\end{figure}

\begin{figure}[htbp]
  \centering
  (a)\includegraphics[width=0.4\textwidth]{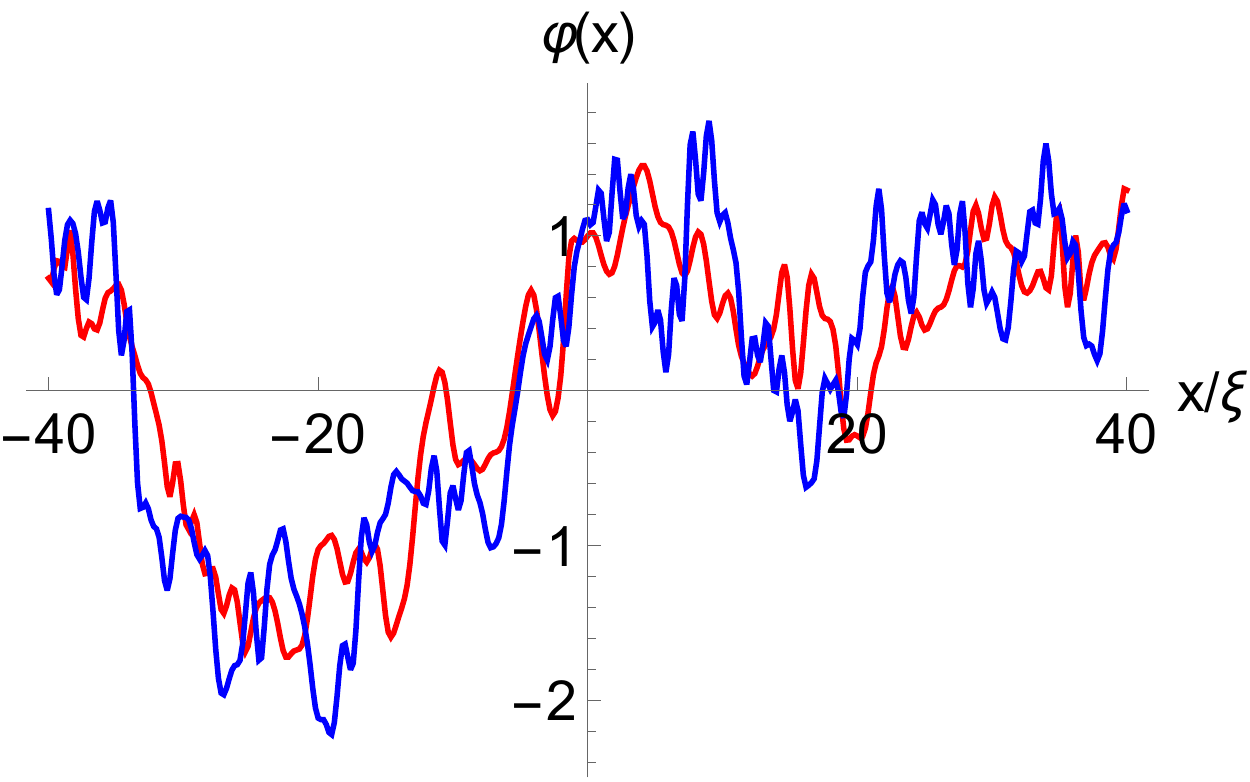}
  \hspace{0.05\textwidth}
  (b)\includegraphics[width=0.4\textwidth]{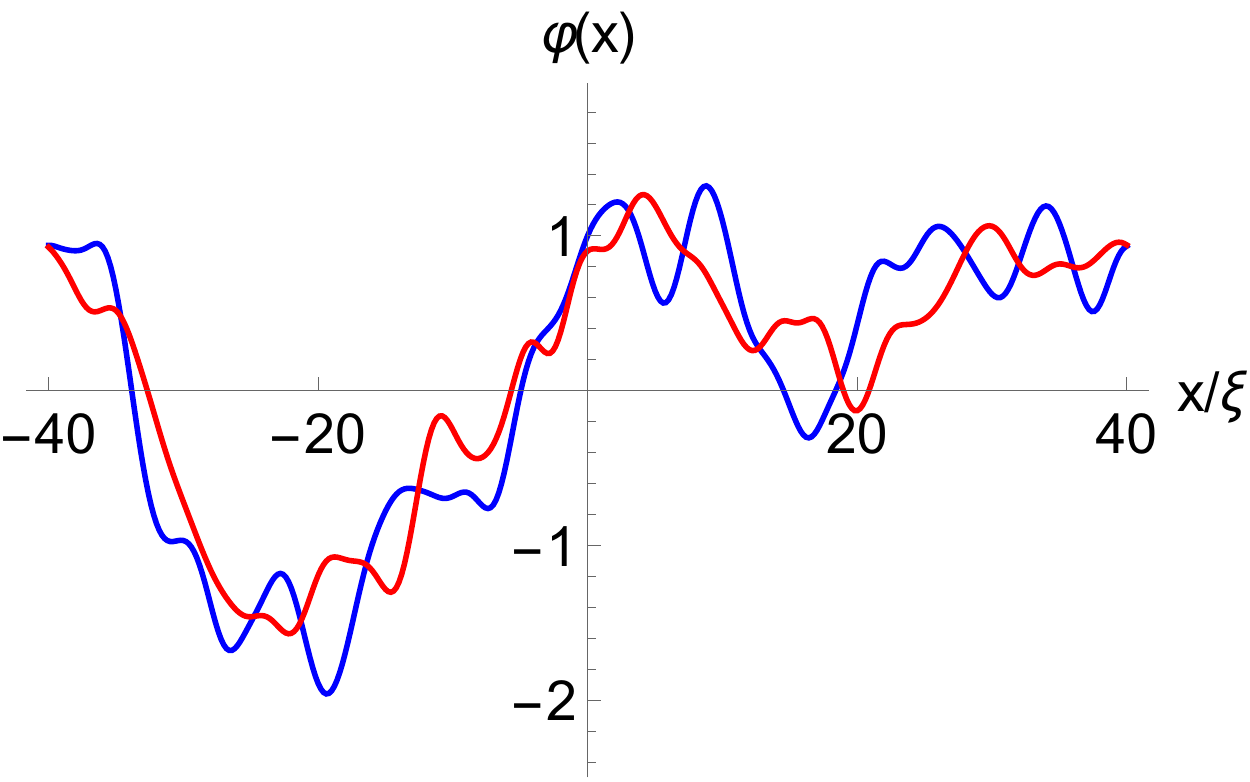}
  \caption{The same as Fig.~\ref{fig:extracted_phases_early}, but at time of flight $t_{1} = 32 \, \mathrm{ms}$, and for a different phase eigenvalue.}
  \label{fig:extracted_phases_late}
\end{figure}

Although the effects of longitudinal expansion included in
(\ref{eq:density_tof_bos_longitudinal_collapsed}) ensure a realistic 
description of the observed gas density, they complicate the
extraction of the eigenvalues $e^{i\varphi_{a}(x,t_0)}$, due to the
presence of $e^{i\varphi_{s}(x,t_0)}$
and the double convolution with a Green's function. Such complications
do not exist for the simplified fit formula (\ref{eigenvalue_simple}),
which neglects longitudinal expansion. This raises the question how
good the results are if, after measuring a density profile given by
(\ref{eq:density_tof_bos_longitudinal_collapsed}), one still
uses eqn (\ref{eigenvalue_simple}) to extract an approximate
eigenvalue $e^{i\tilde{\varphi}_{a}(x,t_0)}$. Having both the full and
approximate expressions at hand, we can explicitly investigate the
accuracy of such an analysis. This is of considerable importance for
the analysis of experiments. To this end, we draw an eigenvalue
$e^{i\varphi_{a}(x,t_0)}$ from the distribution function computed in Section
(\ref{sub:time_dependent_overlap_coefficients}), and construct the
corresponding density profile using
(\ref{eq:density_tof_bos_longitudinal_collapsed}). We then
use the simplified fit formula (\ref{eigenvalue_simple}) to
extract an approximate eigenvalue
$e^{i\tilde{\varphi}_{a}(x,t_0)}$. This can then be compared to the
original, exact eigenvalue
$e^{i\varphi_{a}(x,t_0)}$. Figs~\ref{fig:extracted_phases_early},
\ref{fig:extracted_phases_medium}, \ref{fig:extracted_phases_medium2},
\ref{fig:extracted_phases_late} show representative examples of
such comparisons.

In Fig.~\ref{fig:extracted_phases_early}(a) the extracted phase
$\tilde{\varphi}_{a}(x)$ (red) is compared to the exact phase
$\varphi_{a}(x)$ (blue). Although the results clearly deviate, most of
these deviations occur on small lengthscales, which are not observed
in experiment. To remove these short wave length fluctuations we
convolve the signal with a Gaussian kernel of width $\xi$. The
resulting smoothened curves are seen to be in good agreement for short time
of flight (Fig.~ \ref{fig:extracted_phases_early}, with $t_{1} = 4 \,
\mathrm{ms}$), whereas significant deviations do occur for long flight
times (Fig.~ \ref{fig:extracted_phases_late}, with $t_{1} = 32 \,
\mathrm{ms}$). The size of these deviations does not depend strongly on
the temperature, which only enters through the fields in the symmetric
sector. These symmetric sector fields have an effect on the amplitude 
of the density ripples, but not on the transverse position of the fringes, 
as can be understood by inspection of eqn 
(\ref{eq:density_tof_bos_longitudinal_collapsed}): the eigenvalue $e^{i \varphi_{s}(x)}$ 
appears in both terms in parentheses, so that it does not affect the interference term independently.
For this reason, spatial fluctuations in the eigenvalue $e^{i \varphi_{s}(x)}$ do not strongly
impede the reconstruction of the eigenvalue $e^{i \varphi_{a}(x)}$.


The above analysis leads us to conclude that at sufficiently short
times of flight the simplified fit formula (\ref{eigenvalue_simple})
can be used to obtain an accurate approximation to the eigenvalues
$e^{i\varphi_{a}(x,t_0)}$. 

\begin{figure}[htbp]
\centering
\includegraphics[width=0.32\textwidth]{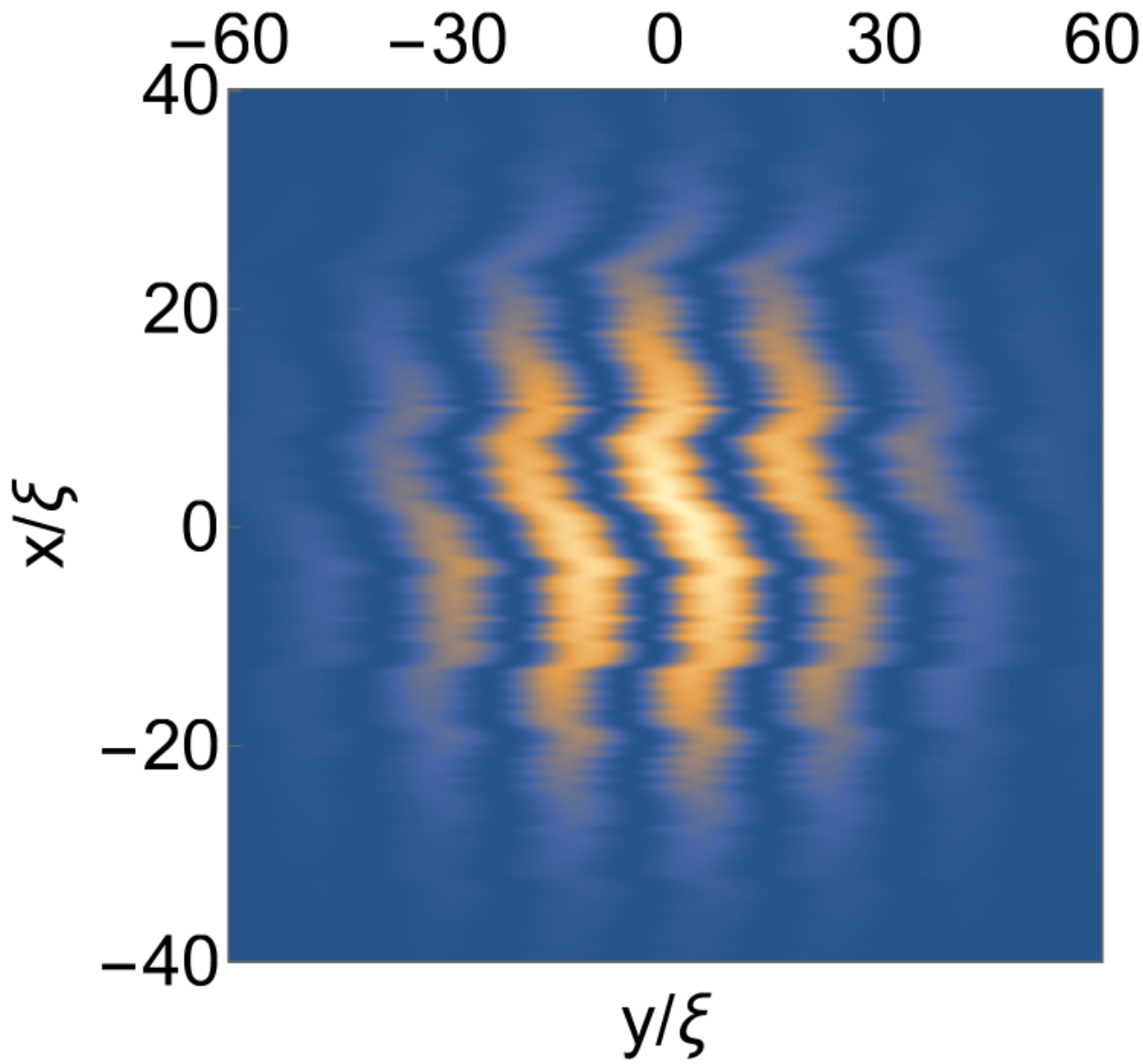}
\includegraphics[width=0.32\textwidth]{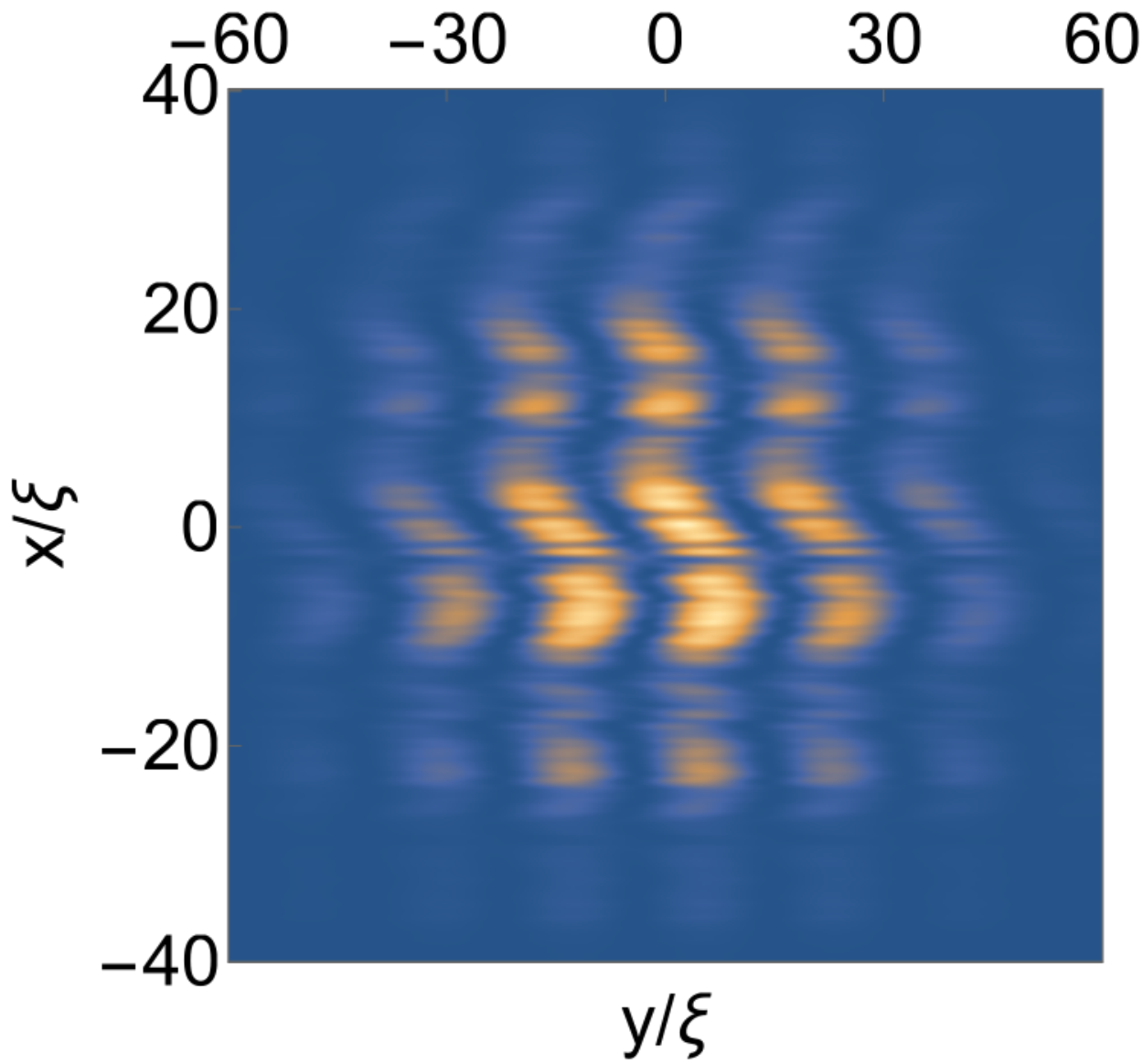}
\hspace{0.5cm}
\includegraphics[trim=435 -63 40 1100,clip,width=0.279\textwidth]{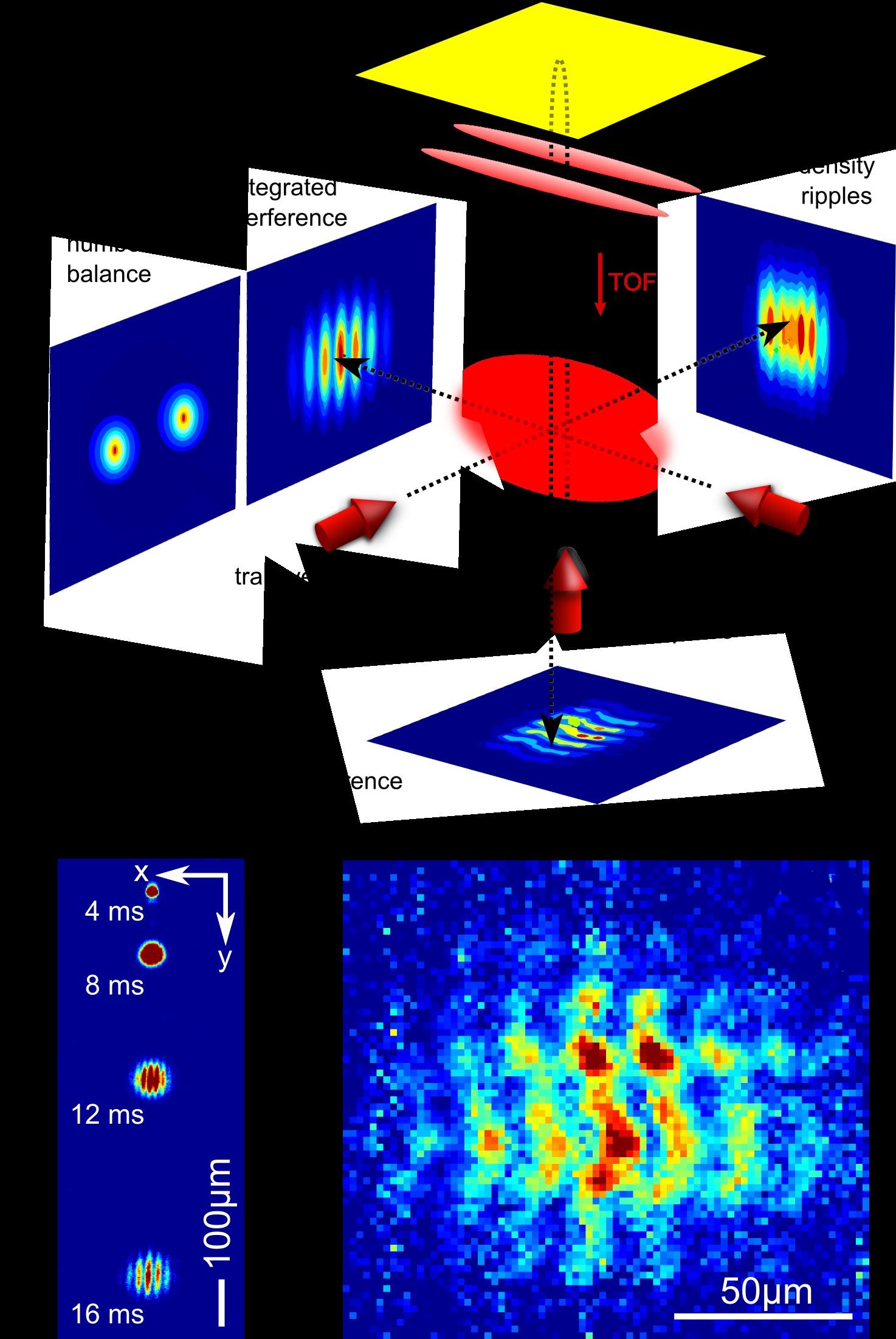}
\caption{(Left) Theoretical results for individual measurement
outcomes $\varrho_{\mathrm{tof}}(\vec{r},t_{1})$ at $t_{0}=14 \,\xi/v$ and 
$t_{1}=16 \,\mathrm{ms}$. An overall suppression with a factor $e^{-
  x^{2}/\left( L/4 \right)^{2} }$ has been applied along the length of
the gas (see main text). Longitudinal expansion during 
time-of-flight has been neglected and the parameters are described
in Section \ref{sub:experimental_parameters}, with $k_{\mathrm{B}}T = 0.5 \, \hbar \omega_{\perp}$, 
so that $T \approx 34 \, \mathrm{nK}$. (Middle) Same as left panel but 
with longitudinal expansion taken into account using
(\ref{eq:density_tof_bos_longitudinal_approx}). 
(Right) Experimentally measured density profile taken from 
Ref.~\cite{Schaff2015}.} 
\label{fig:Dens_experiment}
\end{figure}

In order to compare to experimental data one also should model the
effects of the trapping potential. This can be done in the framework
of a local density approximation \cite{Petrov2000,Petrov2004,Kheruntsyan2005,Citro2008,Geiger2014}. We refrain
from presenting such an analysis here, but instead simply introduce
an overall suppression $e^{-  x^{2}/\left( L/4 \right)^{2} }$ along 
the length of the gas. In Fig.~\ref{fig:Dens_experiment} we present a
comparison of theoretical results obtained in this way to experimental
data from Ref.~\cite{Schaff2015}. We see that the theoretical result
reproduces the various structures seen in experiment. Due to the
statistical nature of measurements in quantum theory the outcome shown
in the theoretical plot is of course not expected to coincide with
that of the experimental plot.

\section{Conclusions} 
\label{sec:conclusions}
In this work we have revisited the theoretical description of the measurement
process involved in time-of-flight recombination of split
one-dimensional Bose gases. We have derived the relation between the
measured density operator after expansion and local operators in
the Luttinger liquid theory  describing the low energy degrees of
freedom in such systems. In the weakly interacting regime and in cases
where the longitudinal expansion can be neglected the measured
density is related in a simple way to a vertex operator of the
phase field in a Luttinger liquid. We have discussed the theoretical
description of individual (projective) measurements in this
setting. To the best of our knowledge this issue has not
been previously addressed in the literature.
We also have described how multi-point correlation
functions of vertex operators can be extracted from projective
measurements of the boson density in time of flight experiments. Our
main new result, which is of direct relevance for experiments, is the
description of projective density measurements in the framework of
Luttinger liquid theory in the case of weak interactions but
non-negligible longitudinal expansion of the gas after the trap
release. Here the main new effect is that phase fluctuations in the
symmetric sector induce intensity variations along the interference
fringes (``density ripples''), the magnitude of which increases with
time of flight. As an explicit example we considered the case of
weakly interacting coherently split Bose gases in the absence of
tunnel coupling. In this case the time evolution can be analyzed
explicitly in the framework of Luttinger liquid theory, see
e.g. \cite{Kitagawa2011}. Our results for a single measurement
reproduce all the main features seen in experiment.  
The theoretical framework developed here applies equally to the case
of weakly interacting split condensates in the presence of a weak tunnel
coupling. Here the antisymmetric sector of the theory is described by
a quantum sine-Gordon model in the weak interaction regime and the
time evolution can no longer be analyzed in a simple fashion.
Our work raises a number of interesting questions. First
and foremost our result \fr{eq:density_tof_bos_longitudinal_collapsed}
suggests that it should be possible to extract information on the
symmetric sector of the theory from the density ripples along the
interference fringes. An investigation of this issue is under
way. Having direct experimental access to properties of the symmetric
sector is important as the existing theoretical analyses suggest that
the relaxational behaviour of the symmetric sector is very different
from that of the  antisymmetric sector.

\paragraph{Acknowledgements}
We are grateful to the Erwin Schr\"odinger International Institute for
Mathematics and Physics for hospitality and support during the
programme on \emph{Quantum Paths}. This work was supported by the EPSRC under grant EP/N01930X (FHLE) and YDvN is supported by the Merton College Buckee Scholarship and the VSB and Muller Foundations. JS acknowledges support by the European Research Council, ERC-AdG \textit{QuantumRelax} (320975). 

\begin{appendix}
\section{Relation between density operators before and after release} 
\label{sec:relation_between_density_operators_before_and_after_release}

We here present the details of the derivation of eqn (\ref{eq:final_relation_field_operator_tof}), by performing the integrals in (\ref{eq:expansion_time_evol_intermed}),
\begin{align}
\hat{\Psi}_{\mathrm{tof}}(x, \vec{r}, t_{1} + t_{0}) = \int \frac{dk\,d^{2}\vec{p}\,dy\,d^{2}\tilde{\vec{r}}}{\left( 2 \pi \right)^{3} } e^{-ik(x-y)}e^{-i \vec{p} \cdot \left( \vec{r} - \tilde{\vec{r}} \right) } e^{-i t_{1} \frac{k^{2}+\vec{p}^{2}}{2m}} \hat{\Psi}(z, \tilde{\vec{r}}, t_{0}),
\end{align}
after insertion of relation (\ref{eq:field_op_two_wells}),
\begin{align}
\hat{\Psi}(x, \vec{r}, t_{0}) = \hat{\psi}_{1}(x, t_{0})g(\vec{r} + \vec{d}/2) + \hat{\psi}_{2}(x, t_{0})g(\vec{r} - \vec{d}/2),
\end{align}
where $g_1(\vec{r})$ is the ground state wave function of a
two-dimensional harmonic oscillator with frequency $\omega$,
\begin{align}
g_{1}(\vec{r}) = \sqrt{\frac{m \omega}{\pi}} e^{-\frac{m \omega}{2} \vec{r}^{2}}.
\end{align}
Defining $\psi_{1} \equiv \psi_{-}$ and $\psi_{2} \equiv \psi_{+}$ and carrying out the integrals we have
\begin{align}
  \hat{\Psi}_{\mathrm{tof}}(x, \vec{r}, t_{1} + t_{0}) &=
 \sum_{\pm} \sqrt{\frac{m \omega}{\pi}} \frac{e^{-i \pi /2} e^{i \arctan \frac{1}{\omega t_{1}}}}{\sqrt{1+\omega^{2} t_{1}^{2}}} \exp \left( - \frac{m \omega}{2} \frac{\left( \vec{r} \pm \vec{d}/2 \right)^{2}}{1+\omega^{2} t_{1}^{2}} \right) \times \nn 
&\times\exp \left( i \frac{m \omega^{2} t_{1}}{2\left(1+\omega^{2} t_{1}^{2}\right)} \left( \vec{r} \pm \vec{d}/2 \right)^{2} \right) \int dy\, G \left( x-y,t_{1} \right) \hat{\psi}_{\pm}(y, t_{0}),
\label{rhogeneral}
\end{align}
where we have defined the free, single-particle Green's function
\begin{align}
G(y,t) = \int \frac{dk}{2 \pi} e^{-iky} e^{-i \frac{t \gamma}{2m} k^{2}} = \begin{cases}
  \sqrt{\frac{m}{2 \pi i t \gamma}} \exp \left( i \frac{m}{2t \gamma} y^{2} \right), &\text{ if }\gamma = 1\\
  \delta(y), &\text{ if }\gamma = 0.
\end{cases} \label{eq:Greens_fn}
\end{align}
We are interested in the limit of a very narrow trapping
potential. Assuming that $\omega t_1\gg 1$ and $|\vec{r}|\gg
|\vec{d}|$ we may simplify \fr{rhogeneral} further, to
\begin{align}
\hat{\Psi}_{\mathrm{tof}}(x, \vec{r}, t_{1} + t_{0}) &\approx -i\sum_{\pm}
\hat{\psi}_{\pm}(x) 
\sqrt{\frac{m\omega}{\pi(1+\omega^{2} t_{1}^{2})}}
\exp \left( - \frac{m \omega}{2} \frac{\vec{r}^{2}}{1+\omega^{2} t_{1}^{2}} \right) \times \nn
&\times \exp \left( i \frac{m}{2t_{1}} \left( \vec{r} \pm \vec{d}/2 \right)^{2} \right) \int dy\, G \left( x-y,t_{1} \right) \hat{\psi}_{\pm}(y, t_{0}).
\end{align}
From this expression, we recover eqn
(\ref{eq:final_relation_field_operator_tof}) with
\begin{align}
f( \vec{r},t_{1}) = -i\sqrt{\frac{m \omega}{\pi}} \frac{1}{\sqrt{1+\omega^{2} t_{1}^{2}}} \exp \left( - \frac{m \omega}{2} \frac{\vec{r}^{2}}{1+\omega^{2} t_{1}^{2}} \right).
\end{align}

\section{Bosonization conventions} 
\label{sec:details_of_the_bosonized_hamiltonian}
The low-energy physics of the microscopic Hamiltonian
(\ref{eq:micr_1d_ham}) is described by a \textit{Luttinger liquid}
\cite{Haldane1981,Cazalilla2004,Kitagawa2011} with Hamiltonian
\begin{align}
H_{\mathrm{LL}} &= \frac{v}{2\pi}\sum_{j=s,a} \int_{-L/2}^{L/2} dx \left[K (\partial_{x} \hat{\phi}_{j}(x))^{2} + \frac{1}{K} \left( \partial_{x} \hat{\theta}_{j}(x) \right)^{2} \right].\label{eq:Luttinger_Liquid_ham}
\end{align}
The (real) fields $\hat{\phi}_{a,s}$ and $\hat{\theta}_{a,s}$ are related to
the original complex bosons $\psi_{1,2}$ by the transformation
(\ref{eq:symm_antisymm_trafo}) and the bosonization identity
\begin{align}
\psi_{j}^{\dagger}(x) \sim \sqrt{\rho_{0} + \frac{\partial_{x} \hat{\theta}_{j}(x)}{\pi}}\; e^{-i \hat{\phi}_{j}(x)} \sum_{m} A_{m} e^{2im \left( \hat{\theta}_{j}(x) + \pi  \rho_{0} x\right)},\;\;\;\; j=1,2. \label{eq:bosonization_identity_app}
\end{align}
Here $A_m$ are non-universal coefficients, $\partial_{x} \hat{\theta}_{1,2}$
describe density fluctuations and $\hat{\phi}_{1,2}$ are phase fields. They
satisfy canonical commutation relations
\begin{align}
    \left[ \frac{\partial_{x} \hat{\theta}_{i}(x)}{\pi}, \hat{\phi}_{j}(z) \right] = i \delta_{i,j} \delta (x - z).
\end{align}
The cutoff length scale for the low-energy field theory
\fr{eq:Luttinger_Liquid_ham} is set by the healing length of the gas,
which  for weak interactions reads $\xi = \pi / mv$.  
The Hamiltonian \fr{eq:Luttinger_Liquid_ham} is parametrized by the
velocity $v$ and the \textit{Luttinger parameter}, $K$. For weak
interactions they are related to the parameters of the microscopic
Hamiltonian (\ref{eq:micr_1d_ham}) as follows \cite{Cazalilla2004}
\begin{align}
v = \frac{\rho_{0}}{m} \sqrt{\gamma} \left( 1 - \frac{\sqrt{\gamma}}{2 \pi}\right)^{1/2}, \;\; K =  \frac{2 \pi}{\sqrt{\gamma}} \left( 1 - \frac{\sqrt{\gamma}}{2 \pi}\right)^{-1/2}, \;\; \rho_{0} = \frac{2mvK}{\pi}, \label{eq:micr_pars}
\end{align}
where we have used the dimensionless parameter $\gamma = m g/\rho_{0}$.

We use periodic boundary conditions throughout this paper. As
$\hat{\phi}_{a,s}$ are compact fields we have
\be
\hat{\phi}_{a,s}(x+L) = \hat{\phi}_{a,s}(x) + 2 \pi \hat{J}_{a,s},
\ee
where the eigenvalues of $\hat{J}_{a,s}$ are integers related to the
number of times the phase winds around a circle of radius $2\pi$ over
the length of the gas. The density operator has to satisfy
\be
\int_{0}^{L} dx \,\partial_{x} \hat{\theta}_{a,s}= \pi \delta
\hat{N}_{a,s}\ ,
\ee
where $\delta \hat{N}_{a,s}$ has integer eigenvalues which count the 
particle imbalance in the symmetric and antisymmetric sectors
respectively. These considerations lead to the mode expansions 
\begin{align}
\hat{\theta}_{j}(x) &= \hat{\theta}_{j,0} +\frac{\pi x}{L}\delta \hat{N}_{j} + \sum_{q\neq 0} \Big| \frac{\pi K}{2 q L} \Big|^{1/2} e^{iqx}  \left(\hat{a}_{j,q} + \hat{a}^{\dagger}_{j,-q} \right), \label{eq:mode_exp_dens_caz}\\
\hat{\phi}_{j}(x) &= \hat{\phi}_{j,0} + \frac{\pi x}{L} \hat{J}_{j} + \sum_{q\neq 0} \Big| \frac{\pi}{2q L K} \Big|^{1/2} \mathrm{sgn}(q) e^{iqx} \left( \hat{a}_{j,q} - \hat{a}^{\dagger}_{j,-q} \right), \label{eq:mode_exp_phase_caz}
\end{align}
where $\hat{a}_{i,q}$ are oscillator modes with commutation relations $[
\hat{a}_{i,q},\hat{a}^{\dagger}_{j,k}]=\delta_{q,k}\delta_{i,j}$, and
$[ \delta \hat{N},   \hat{\phi}_{0} ]=i=[ \hat{J}, \hat{\theta}_{0} ]$. The momenta 
are quantized as $q_{n} = 2 \pi n/L$. The mode expansion of the
Hamiltonian (\ref{eq:Luttinger_Liquid_ham}) is 
\begin{align}
H_{\mathrm{LL}} = \sum_{j = a,s} \left[ \frac{\pi v K \hat{J}_{j}^{2}}{2L} + \frac{\pi v (\delta \hat{N}_{j})^{2}}{2KL} + \sum_{q \neq 0} v |q| \hat{a}_{j,q}^{\dagger} \hat{a}_{j,q} \right] . 
\end{align}
For our purposes it will suffice to consider only the $\hat{J}=0$
subspace. The rationale for this is that $\hat{J}$ has eigenvalue zero
for all experimentally relevant initial states and the Hamiltonians we
consider commute with $\hat{J}$.

A compact notation for the zero modes used in eqns
(\ref{eq:phi_vec_form_full}-\ref{eq:del_theta_vec_form_full}) is to
introduce annihilation operators
\begin{align}
\hat{a}_{a,0} = - i\sqrt{\frac{2K}{v}} \hat{\phi}_{a,0}-
\frac{1}{2}\sqrt{\frac{v}{2K}} \delta \hat{N}_{a}\ .
\label{eq:ladder_op_zm}
\end{align}

\section{Normalization of vertex operator eigenstates} 
\label{sec:normalization_of_vertex_operator_eigenstates}
We here derive eqns (\ref{eq:normalization}) and
(\ref{eq:delta_normalized}). In order to regulate the infinity caused
by the delta function, we consider the following modification of the state
(\ref{eq:eigenstate})
\begin{align}
\ket{\{f_{n}\}}_{\tau} = \mathcal{N}_{f} \exp \sum_{k} \left( \frac{\tau}{2} \hat{a}_{k}^{\dagger}\hat{a}_{-k}^{\dagger} + \frac{f_{k}}{u_{k}} \hat{a}_{k}^{\dagger} \right) \ket{0},
\end{align}
and recover the eventual delta function normalization by taking the
limit $\tau \rightarrow 1$ at the end of the calculation. Our task is
to calculate the overlap
\bea
\langle g|f\rangle&=&_{\tau}\braket{\{g_{n}\}|\{f_{n}\}}_{\tau}\nn
&=& \mathcal{N}_{g}^{*}\mathcal{N}_{f} \bra{0} \exp \left(\sum_{j} \frac{\tau}{2}\hat{a}_{j}\hat{a}_{-j} + \frac{g_{j}^{*}}{u_{j}^{*}} \hat{a}_{j} \right) \exp \left(\sum_{k} \frac{\tau}{2} \hat{a}_{k}^{\dagger}\hat{a}_{-k}^{\dagger} + \frac{f_{k}}{u_{k}} \hat{a}_{k}^{\dagger} \right)  \ket{0}.
\eea
Inserting a resolution of the identity in terms of normalized coherent
states 
\begin{align}
    \ket{\alpha} = \prod_{k}e^{- |\alpha_{k}|^{2}/2} e^{\alpha_{k}^{\vphantom{\dagger}} a_{k}^{\dagger}} \ket{0},\qquad
\mathbbm{1}=\int D\left(\alpha,\alpha^{*}\right) \ket{\alpha}\bra{\alpha}
\end{align}
with $D\left(\alpha,\alpha^{*}\right)
\ket{\alpha}\bra{\alpha}=\prod_{k}d {\rm Re}\alpha_{k}\;d{\rm Im}\alpha_{k}$
and using that $a_{k}\ket{\alpha} = \alpha_{k}\ket{\alpha}$ we have
\begin{align}
\langle g|f\rangle = \int D\left(\alpha,\alpha^{*}\right) \mathcal{N}_{g}^{*}\mathcal{N}_{f} \exp \sum_{j} \left( - \alpha_{j}\alpha^{*}_{j}+ \frac{\tau}{2}\alpha_{j}\alpha_{-j} + \frac{f^{*}_{j}}{u^{*}_{j}} \alpha_{j} + \frac{\tau}{2} \alpha_{j}^{*}\alpha_{-j}^{*} + \frac{f_{j}}{u_{j}} \alpha_{k}^{*} \right).
\end{align}
Noting that $u_{j}$ satisfies
\begin{align}
\begin{cases}
	{\rm Im} (u_{j}) = 0 , \;\;\;\;u_{-j} = -u_{j}, &\text{ if } j \neq 0,\\
	{\rm Re} (u_{0}) = 0, &\text{ else, }
\end{cases}
\end{align}
and using $f^{*}_{-n}=f_{n}$ and $f_{0}^{*}=f_{0}$ we can carry out
the integrals. Finally we use that
\begin{align}
\lim_{\epsilon \rightarrow 0} \frac{1}{\left(2 \pi \epsilon\right)^{d/2}}e^{-\frac{|x|^{2}}{2 \epsilon}} = \delta^{(d)} \left( |x| \right)
\end{align}
to arrive at
\begin{align}
\begin{split}
\lim_{\tau \rightarrow 1}\braket{\{g_{n}\}|\{f_{n}\}}_{\tau} = \mathcal{N}_{g}^{*}\mathcal{N}_{f} \sqrt{2\pi} |u_{0}| &\exp \left( \frac{1}{8|u_{0}|^{2}}\left( g_{0} + f_{0} \right)^{2} \right) \delta(g_{0} - f_{0})\times \\
\times \prod_{k>0} \pi |u_{k}|^{2} &\exp \left( \frac{1}{4|u_{k}|^{2}} \left| g_{k} + f_{k} \right|^{2} \right) \delta^{(2)}\left(  g_{k} - f_{k} \right).
\end{split}
\end{align}
This shows that the states $\ket{\{f_{n}\}}$ are delta-normalized if
the normalization constants $\mathcal{N}_{f}$ are chosen according to
eqn (\ref{eq:normalization}). 

\section{Time-dependent overlap for the zero mode} 
\label{sec:time_dependent_overlap_for_the_zero_mode}


The zero mode initial state $\ket{\psi_{k=0}}$ is determined by the overlap
\begin{align}
\braket{n | \psi_{k=0}} = \left( \frac{1}{\pi \rho_{0} L} \right)^{1/4} \exp \left( - \frac{1}{2 \rho_{0} L} n^{2} \right),
\end{align}
where $\ket{n}$ is the eigenstate of $\delta \hat{N}$ with eigenvalue $n$. The operators $\delta \hat{N}$ and $\hat{\phi}_{0}$ satisfy canonical commutation relations, $\left[ \delta \hat{N}, \hat{\phi}_{0} \right] = i$. In analogy with eigenstates of the $\hat{x}$- and $\hat{p}$-operators in quantum mechanics, this means that the eigenstate $\ket{f_{0}}$ of $\hat{\phi}_{0}$ has an overlap with the eigenstate $\ket{n}$ of $\delta \hat{N}$ which is given by
\begin{align}
\braket{n | f_{0}} = \frac{e^{i n f_{0}}}{\sqrt{2 \pi}}.
\end{align}
We are interested in computing the time-dependent overlap
\begin{align}
\braket{f_{0}|\psi_{k=0}(t)} = \bra{f_{0}} e^{-i H_{k=0}t}\ket{\psi_{k=0}}.
\end{align}
Since the zero mode part of the Hamiltonian is given by
\begin{align}
H_{k=0} = \frac{\pi v (\delta \hat{N})^{2}}{2KL},
\end{align}
its action on the state $\ket{n}$ is trivial, and we can compute the time-dependent overlap by inserting a complete set of such states. This leads to the result that
\begin{align}
\bra{f_{0}} e^{-i H_{k=0}t}\ket{\psi_{k=0}} &= \int dn \, \braket{f_{0}|n} \bra{n} e^{-i H_{k=0}t}\ket{\psi_{k=0}} \nn
&= \left( \frac{1}{\pi \rho_{0} L} \right)^{1/4} \frac{1}{\sqrt{\frac{1}{\rho_{0} L} + i \frac{\pi v t}{K L}}} \exp \left( - \frac{1}{2} \frac{f_{0}^{2}}{\frac{1}{\rho_{0} L} + i \frac{\pi v t}{K L}} \right) 
\end{align}

\section{Overlap with a general Fock state} 
\label{app:overlap_fock_state}

We here compute the overlaps between a generic phase eigenstate (\ref{eq:eigenstate}) and a Fock state $\ket{\{n_{q \neq 0}\}}$, where we assume that the occupation numbers satisfy $n_{q} = n_{-q}$. The zero mode will not be treated here. Defining 
\begin{align}
\mathcal{N}_{q} = \left(\frac{1}{\pi |u_{q}|^{2}}\right)^{1/2} e^{- \frac{1}{2|u_{q}|^{2}} |f_{q}|^{2}}, 
\end{align}
we consider sectors $(q,-q)$ separately. This leads to
\begin{align}{}
  \braket{n_{-q},n_{q}|f_{-q},f_{q}} &= \mathcal{N}_{q}\bra{n_{-q},n_{q}} \sum_{n=0}^{\infty} \frac{1}{n!} \left( \hat{a}^{\dagger}_{q} \hat{a}^{\dagger}_{-q} + \frac{f_{q}}{u_{q}} \hat{a}_{q}^{\dagger} + \frac{f_{q}^{*}}{u_{q}^{*}} \hat{a}_{-q}^{\dagger} \right)^{n} \ket{0} \nn
  &= \mathcal{N}_{q} n_{q}! \sum_{n=0}^{\infty} \frac{1}{n!} \left( \frac{f_{q}}{u_{q}} \right)^{\alpha} \left( \frac{f_{q}^{*}}{u_{q}^{*}} \right)^{\gamma} C(\alpha,\gamma),
\end{align}
with
\begin{align}
\alpha &= n- n_{q} = \gamma,
\end{align}
and $n_{q} \leq n \leq 2n_{q}$. The combinatoric factors read
\begin{align}
C(\alpha,\gamma) = \begin{pmatrix}
  n \\
  \alpha + \gamma
\end{pmatrix} \begin{pmatrix}
  \alpha + \gamma \\
  \gamma
\end{pmatrix} = \frac{n!}{(2n_{q} - n)!((n-n_{q})!)^{2}}.
\end{align}
The overlap in the $(q,-q)$-sector is then given by
\begin{align}
\braket{n_{-q},n_{q}|f_{-q},f_{q}} &= \mathcal{N}_{q} \sum_{n=n_{q}}^{2n_{q}} \frac{n_{q}!}{(2n_{q} - n)! \left( \left( n-n_{q} \right) ! \right)^{2} } \left( -1 \right)^{n - n_{q}} \Big| \frac{f_{q}}{u_{q}} \Big|^{2n-2n_{q}} \nn
&= \mathcal{N}_{q} \, L_{n_{q}} \left( \Big| \frac{f_{q}}{u_{q}} \Big|^2 \right),
\end{align}
where $L_{n}(x)$ is the Laguerre polynomial of degree $n$. Inserting the definition of $\mathcal{N}_{q}$, we find the squared overlap coefficients per $(q,-q)$-sector,
\begin{align}
\big| \braket{n_{-q},n_{q}|f_{-q},f_{q}} \big|^{2} = \frac{1}{\pi |u_{q}|^{2}} L^{2}_{n_{q}}\left( \Big| \frac{f_{q}}{u_{q}} \Big|^2 \right) e^{- \big| \frac{f_{q}}{u_{q}} \big|^2}.
\end{align}

\end{appendix}

\end{document}